\documentclass[]{./Forme/jfm}

\usepackage{graphicx}
\usepackage{tikz}
\usepackage{xcolor}
\usepackage{newtxtext}
\usepackage{newtxmath}
\usepackage{amsmath}
\usepackage{natbib}
\usepackage[backref = page]{hyperref}
\usepackage{physics}
\usepackage{subcaption}
\usepackage{caption} 
\usepackage{float}      
\usepackage{lipsum}
\usepackage[export]{adjustbox} 

\usetikzlibrary{tikzmark,calc} 

\usepackage{tikz}
\usetikzlibrary{shapes,arrows}
\tikzstyle{decision} = [diamond, draw, fill=blue!20, 
    text width=4.5em, text badly centered, node distance=3cm, inner sep=0pt]
\tikzstyle{block} = [rectangle, draw, fill=re!20, 
    text width=5em, text centered, rounded corners, minimum height=4em]
\tikzstyle{line} = [draw, -latex']
\tikzstyle{cloud} = [draw, ellipse,fill=red!20, node distance=3cm,
    minimum height=2em]

\hypersetup{
    colorlinks = true,
    urlcolor   = blue,
    citecolor  = blue,
}

\newcommand{\RomanNumeralCaps}[1]
\linenumbers

\usepackage{./Forme/math_variables_and_commands}

\title{Receptivity and instability of the hypersonic flow over moderately blunt cones}

\author{N. d'Eprémesnil\aff{1}
  \corresp{\email{nicolas.du-val-d-epremesnil@ensma.fr}},
  C. Caillaud \aff{1} \corresp{\email{clement.caillaud@cea.fr}}, G. Lehnasch \aff{2}\corresp{\email{guillaume.lehnasch@ensma.fr}}, M. Olazabal \aff{1} \corresp{\email{marina.olazabal@ensma.fr}}
 \and P. Jordan \aff{2} \corresp{\email{peter.jordan@univ-poitiers.fr}}}
\author{N. d'Eprémesnil\aff{1}
  \corresp{\email{nicolas.du-val-d-epremesnil@ensma.fr}},
  C. Caillaud \aff{1}, G. Lehnasch \aff{2}, M.Olazabal \aff{1} 
 \and P. Jordan \aff{2} }

\affiliation{\aff{1}CEA-CESTA, 15 Avenue des Sablières, CS60001, 33116 Le Barp Cedex, France
\aff{2}Département Fluides, Thermique, Combustion, Institut PPRIME, CNRS – Université de Poitiers – ENSMA, UPR 3346, 86036 Poitiers, France}

\begin{document}

\maketitle

\begin{abstract}

With a view to identifying and understanding the linear receptivity and amplification mechanisms that underpin laminar-to-turbulent transition over blunt bodies in hypersonic flow, we use resolvent analysis to study the flow over a blunt cone with $7^\circ$ half-angle at Mach number $\M_\infty = 6$, zero angle of attack, and nose-radius–based Reynolds number $\Rey_{R_n} = 90000$. Optimal forcing and responses are obtained for frequencies up to \SI{330}{\kHz} and azimuthal wavenumbers between 0 and 200. Wall-temperature effects are accounted for by considering both isothermal ($T_w =$ \SI{300}{\K}) and adiabatic wall conditions.
The resolvent analysis shows that stationary streak modes are the most amplified in the isothermal case, followed by entropy-layer modes between $20$ and $140~\SI{}{\kHz}$. In the adiabatic case, the $1^{st}$ Mack mode is the most amplified. 
The entropy layer, caused by the nose-tip bluntness, has a profound influence on the receptivity structures. For the optimal streak mode, the most intense receptivity structures lie deep in the entropy layer, further away from the boundary-layer edge compared to the equivalent sharp-cone streak mode. This indicates that atmospheric disturbances may excite streak-like instability without fully penetrating the boundary layer.
For the entropy-layer modes, the dominant receptivity and fluctuation signatures are located within the entropy layer. An energy-budget analysis reveals that these modes are most susceptible to kinetic disturbances and they sign the most on temperature fluctuations. These modes are found to leverage a temperature mixing mechanism that exploits the baseflow’s wall-normal temperature gradient in the entropy layer to grow.

\end{abstract}


\hypersetup{linkcolor=red}

\section{Introduction}

Hypersonic flight presents complex vehicle-design challenges.
At these high-speed hypersonic flight conditions, a large amount of kinetic energy of the air flow is converted into heat which is transferred to the vehicle walls.
To sustain the intense heating, most hypersonic vehicles possess both a thermal protection system (TPS) and some bluntness at the tip.
In these conditions the state of the boundary layer, i.e. laminar or turbulent, has an important effect on the heat flux directed into the vehicle. 
Accurate prediction of the position of the laminar-to-turbulent transition is therefore crucial for the design optimisation of hypersonic vehicles. Poor estimation may result in either loss of the vehicle or, in the case of over-cautious design, significant additional weight associated with the heat shield \citep{defense_science_board__washington_dc_report_1988}.
The mechanisms underpinning transition on blunt hypersonic bodies are complex and remain poorly understood, and it is this observation that motivates the present study.

\begin{figure}
\centering
\includegraphics[]{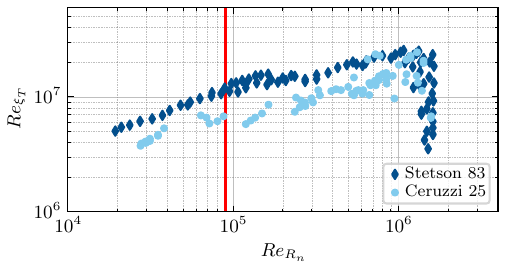}
\captionsetup{width=\linewidth}
\caption[]{Reynolds number of transition versus nose radius based Reynolds number adapted from data from \cite{ceruzzi_observations_2025, stetson_laminar_1983}. The red line is $\Rex{R_n}$ of this study.}
\label{fig:Resultats/paredes_&_ceruzzi_transition}
\end{figure}

In the canonical sharp-cone case, at $0^\circ$ of attack, it is understood that Mack modes are responsible for transition \citep{mack_stability_1987}. 
These are modal, convective instabilities with an exponential spatial growth rate. 
And traditional engineering models such as the $e^N$ method provide satisfactory results for estimation of the transition location on canonical geometries, based on theoretical results of locally parallel Linear Stability Theory (LST). 
However, when bluntness increases, the Mack modes are damped and three flow-transition regimes, with respect to the nose radius, have been identified by \cite{stetson_laminar_1983}:  \emph{small}, \emph{moderate} and \emph{large} bluntness regimes. These findings have been confirmed in recent experiments \citep{marineau_mach_2014, jewell_transition_2018, ceruzzi_experimental_2024} and the transition Reynolds number is plotted in Figure \ref{fig:Resultats/paredes_&_ceruzzi_transition}.
In the \emph{small} bluntness case, LST predicts the transition delay due to the attenuation of Mack modes with increasing bluntness \citep{rosenboom_influence_1999}.
In the \emph{moderate} regime, when bluntness increases further, the baseflow becomes significantly different from the sharp canonical case. 
The blunt nosetip introduces a detached bow shock which creates continuously changing jump conditions along the shockwave. 
This induces a non-uniform shock layer and an entropy layer (EL) appears above the boundary layer (BL). 
In this case, the transition Reynolds number becomes larger and the Mack modes are no longer responsible for transition, which occurs before the Mack-mode amplitude is large enough to cause transition \citep{lei_linear_2012, marineau_mach_2014, paredes_nose-tip_2019}.
As bluntness becomes even more important, the \emph{large} bluntness regime is reached. 
In this regime, the transition Reynolds number suddenly reduces, see $\Rex{R_n} > 10^6$ in Figure \ref{fig:Resultats/paredes_&_ceruzzi_transition}. 
This phenomenon is referred to as \emph{transition reversal}. 
\cite{reshotko_role_2004} and \cite{paredes_blunt-body_2017} argue that in this regime, transient growth due to nosetip surface roughness could explain the decrease in transition Reynolds number.

The mechanisms responsible for transition, in the \emph{moderate} and \emph{large} bluntness regimes are not yet well understood. 
We thus consider linear amplification mechanisms that may arise in the \emph{moderate} bluntness regime. 
Recent studies have shown that non-modal growth may explain transition  \citep{paredes_nonmodal_2019, paredes_mechanism_2020, cook_understanding_2018}.  
Optimal growth theory explored in the framework of the Harmonic Linearised Navier-Stokes Equations (HLNSE) and the Parabolised Stability Equations (PSE) has shown that substantial transient growth of streak and entropy layer modes modes may occur. 
\cite{paredes_mechanism_2020} later performed Direct Numerical Simulation (DNS) at $\M_\infty = 5.9$ and $\Re_{R_n} = 139 000$, where a pair of optimal perturbations at \SI{250}{\kHz}, corresponding to oblique entropy layer disturbances, were shown to produce transition through the generation of stationary streaks (via a triadic interaction of the oblique entropy layer modes) in the boundary layer.
Though the initial perturbation was located in the entropy layer, the non-linear interaction led to an excitation of the boundary layer.
At similar conditions, linear DNS shows larger growth associated with entropy layer perturbation than that captured by PSE analysis \citep{hartman_nonlinear_2021} indicating that optimal PSE may miss the truly optimal linear mechanisms. 
Recently, schlieren visualisation \citep{grossir_influence_2019, kennedy_visualizations_2019, kennedy_characterization_2022, ceruzzi_experimental_2024}, has shown elongated wisp-like structures in the entropy layer that are qualitatively consistent with entropy layer modes found in linear analysis. This further supports the scenario of transient growth underpinned by entropy layer mechanisms in the \emph{moderate} bluntness regime. 

While entropy layer modes appear to be relevant for the transition process, it is important to explore in a systematic fashion all possible linear amplification mechanisms. 
To date, there has not been such a systematic study of the receptivity and response of the flow for blunt-cone geometries in this regime.
Furthermore, regarding the entropy layer modes, the underlying physical mechanisms are still not well understood, and it remains unclear what type of receptivity leads to their excitation.

To address these questions, we perform a comprehensive investigation of optimal linear amplification mechanisms in hypersonic flow over a blunt conical body in the moderate bluntness regime using global resolvent analysis.
It is performed over a wide range of frequencies and azimuthal wavenumbers, enabling the identification and study of optimal receptivity and amplification mechanisms. 
This provides an upper bound on the amplification of external perturbations via linear processes and permits identification of the  mechanisms that may be triggered by environmental disturbances.
To further understand the underlying physics, a detailed analysis of the dominant optimal modes is undertaken : their energy content and the associated production terms are presented and discussed.

Additionally, during flight, the wall temperature may vary between cold to adiabatic and hot wall conditions. 
Past research has shown that modification of the wall temperature with regard to the recovery temperature can significantly affect the linear amplification mechanisms \citep{mack_stability_1987}. 
Furthermore, flight enthalpy and pressure conditions are difficult to reach in wind tunnel tests and the run durations are often short in front of the model thermal inertia. 
Hence, most hypersonic wind tunnel flows usually only allow for a small range of cold-isothermal wall temperature conditions. 
Therefore, considering that previous numerical stability analysis conducted on the blunt cone have focused solely on such cold isothermal wall conditions, we choose to include an adiabatic wall condition in order to provide insights into the wall temperature effect on blunt cone flow instability. 

The outline of the paper is as follows: Section $\S$\ref{section:Theory} introduces the flow conditions, the geometry and the governing equations of this problem and the associated numerical framework. 
Section $\S$\ref{section:Baseflow_charateristics} presents the baseflow used for the resolvent analysis.
Section $\S$\ref{section:isothermal_resolvent_gain_maps} presents the optimal and suboptimal modes obtained through global resolvent analysis. Two families of modes stand out and are further studied in the next sections. 
In section $\S$\ref{section:OptimalStreakMode}, the optimal streak mode's energy content and production is analysed and in section $\S$\ref{section:entropy_layer_mode_iso} a similar study is conducted on entropy layer modes.
Finally, section $\S$\ref{section:wall_temp_effects} discusses the effect of wall temperature through the resolvent analysis of a baseflow with an adiabatic wall.
Section $\S$\ref{section:Conclusion} concludes and offers perspectives for future research.
\section{Numerical setup and methods}
\label{section:Theory}
\subsection{Geometry and flow conditions}
\label{section:Geometry}
\begin{figure}
\centering
\includegraphics[width = 0.9\linewidth, trim={0.9cm 1.2cm 1.5cm 2cm}, clip]{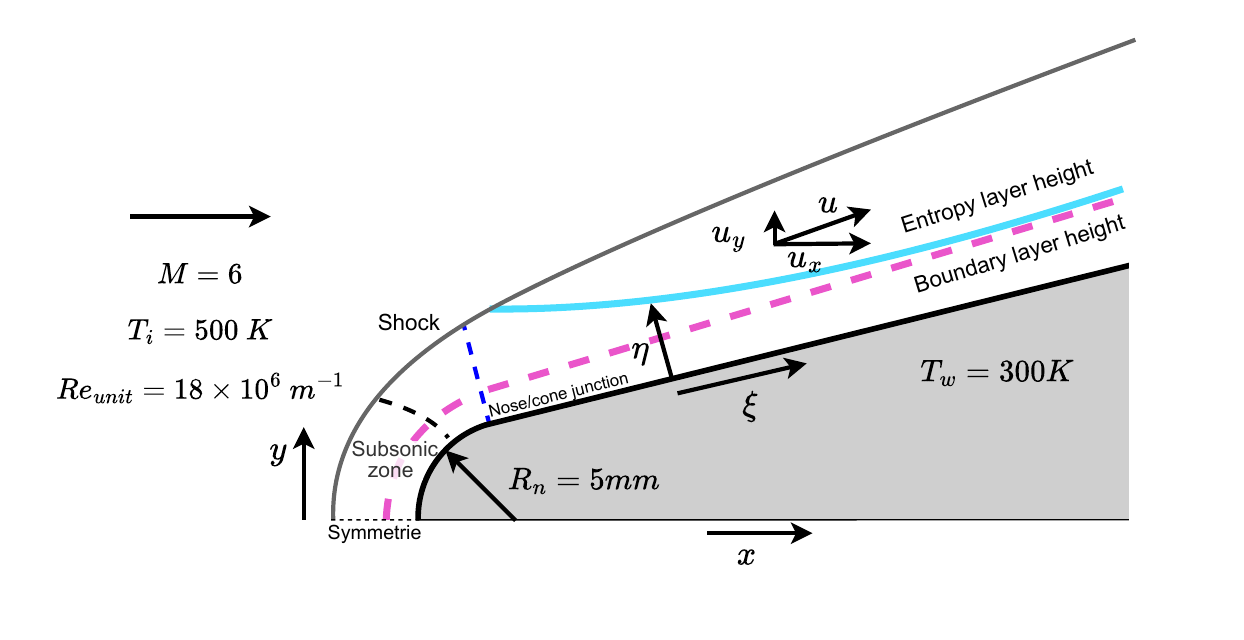}
\captionsetup{width=\linewidth}
\caption[]{Scheme of the flow topology and setup}
\label{fig:Resultats/schema_bf}
\end{figure}

We consider a blunt cone with a spherical nose radius $R_n$, with a conical frustum of half-angle $\alpha =~$\SI{7}{\degree} and \SI{0}{\degree} angle of attack at $\M = 6$. 
The cone is \SI{50}{\cm} long and is truncated at the nose by the distance necessary to fit a spherical nose of radius $R_n$. Figure \ref{fig:Resultats/schema_bf} presents a schematic version of the baseflow and geometry, where $\xi^*$ and $\eta^*$ are respectively the curvilinear distance from the tip and the wall-normal distance.
There are 3 cases considered, summarised in table \ref{Table:studied_configuration}. 

The \SI{5}{\mm} nose radius case with an isothermal wall condition corresponds closely to the experiment B4 of \cite{kennedy_characterization_2022} performed in the AFRL Mach 6 Ludwieg tube. These conditions are transitional, and through similarity based on experiments done by \cite{jewell_transition_2018} in the same wind tunnel, transition occurs at a distance of approximatively \SI{46}{\cm} from the tip.
The B4 experiments performed by \cite{kennedy_characterization_2022} show through high-speed schlieren visualisation the lack of 2nd Mack mode amplification and elongated wisp-like structures in the entropy layer.

\begin{table}
\centering
\begin{tabular}{ccccccc}
name & $R_n \left[\SI{}{\mm}\right]$ & $Re_{R_n}$ & $Re_{\text{unit}}\left[\SI{}{\per\meter}\right]$ &
$T_{i, \infty}  \left[\SI{}{\Kelvin}\right]$ & $M_{\infty}$ & wall boundary condition\\ \hline
sharp isothermal & 0.1 & 1 800 & $18 \times 10^6$ & 500 & 6 & isothermal $T_w = 300$ \SI{}{\Kelvin} \\
blunt isothermal & 5 & 90 000 & $18 \times 10^6$ & 500 & 6 & isothermal $T_w = 300$ \SI{}{\Kelvin} \\
blunt adiabatic & 5 & 90 000 & $18 \times 10^6$ & 500 & 6 & adiabatic $\partial \Temp/\partial \eta |_{wall} = 0$ \SI{}{\Kelvin\per\meter} \\
\end{tabular}
\caption{Studied configurations}
\label{Table:studied_configuration}
\end{table}

\subsection{Governing equations}

The flow is governed by the compressible Navier-Stokes equations, which are non-dimensionalised as follow:
\begin{align}
     {x} &= \frac{x^*}{L_{ref}} ,&  {\Density} &= \frac{\Density^*}{\Density_{\infty}},
    &  {p} &= \frac{p^*}{\Density_\infty \Vvec_\infty^2}, &  {\Vvec} &= \frac{\Vvec^*}{\Vel_\infty}, \nonumber
\end{align}
\begin{align}
     {\Temp} &= \frac{\Temp^*}{\Temp_\infty}, &  {\EnergyStagnationDensity} &= \frac{(\EnergyStagnationDensity)^*}{\Density_\infty \Vel_\infty^2},
    &  {t} &= \frac{t^*\Vel_\infty}{L_{ref}}, \\ 
    \Pran &= \frac{\mu_\infty c_p}{\lambda_\infty}, & \M &= \frac{\Vel^*}{a_\infty}, & \Rey &= \frac{\Density_\infty \Vel_\infty L_{ref}}{\mu_\infty},  \nonumber
\end{align}

\noindent where $\begin{bmatrix} \Density & \Vvec & p \end{bmatrix}^T$ are the usual primitive variables with $\Vvec = (u_x, u_y, u_z)^T$ the Cartesian velocity vector, $\Temp$ the static temperature,  $E = c_v \Temp + \frac{1}{2}\Vvec^2$ and $\lambda$ the thermal conductivity, $\mu$ the dynamic viscosity which follows Sutherland's law, $c_p$ and $c_v$ the specific heat capacities and the subscript $\infty$ refers to unperturbed far field flow conditions and superscript $*$ denotes dimensionalised variables. 
$L_{ref}$ is chosen as the nose radius ($R_n$) of the corresponding blunt cone.
The compressible Navier-Stokes equations are used and can be written compactly as: 
\begin{align}
    \drondt{\q} &=  \Navierq{\q} + \f,  &
    \q &= \begin{bmatrix}
          \Density & \Density\Vvec & \EnergyStagnationDensity
         \end{bmatrix}^T,
         \label{eq:State_of_the_art:Navier_stokes_W_forcing_state_vector}
\end{align}

\noindent where $\Navier$ represents the Navier-Stokes operator, with the viscous stress tensor $\pmb{\tau} = \mu \left( \nabla \Vvec + \left(\nabla \Vvec \right) ^T - \frac{2}{3} \left( \nabla \cdot \Vvec \right) \mathsfbi{I} \right)$, $\mathsfbi{I}$ the identity matrix, $\q$ the conservative-variable vector and $\f$ an external forcing modelling freestream turbulence effects, wind tunnel noise or any other disturbance source terms. A thermally and calorically perfect gas is considered with specific heat ratio $\gamma = 1.4$ and constant Prandtl number $\Pr = 0.72$,
\begin{align}
    p = \frac{\Density\Temp}{\gamma \M_\infty^2}, && ~ c_v = \frac{1}{\gamma \M_\infty^2 \left( \gamma -1 \right)}.
\end{align}

\subsection{Resolvent framework}

As shown by past studies, investigating transition on blunt cones requires to account for the full, modal and non-modal, linear dynamics of the Navier-Stokes operator. Resolvent analysis is well suited for this, as it captures the global linear dynamics and flow receptivity to external perturbations. 
To investigate the linear response of the flow, the Navier–Stokes equations are linearized around a steady baseflow, fixed point of the non-linear system, i.e., $\Navierq{\mean{\q}} = 0$. This is done by expanding the flow state in a first-order Taylor series in $\varepsilon$, where $\varepsilon$ is small enough to justify the linear approximation:
 \begin{align}
    \q = \mean{\q} + \varepsilon \fluctuation{\q}, && 
    ~ \f &= \mean{\f} + \varepsilon  \fluctuation{\f} = \mathbf{0} + \varepsilon \fluctuation{\f},\quad \varepsilon \ll 1,
\end{align}
\begin{align}
    \pdv{\fluctuation{\q}}{t} &= \mathsfbi{J} \fluctuation{\q} + \fluctuation{\f}, \label{eq:Theory:reduced_fluct_NS_system}
\end{align}

\noindent with $\mathsfbi{J} = \left. \pdv{\Navierq{\q}}{\q}\right|_{\mean{\q}}$, the Jacobian computed for the baseflow $\bar{\pmb q}$. To study the flow receptivity and instability we look for the response $\fluctuation{\q}$, associated with the forcing $\fluctuation{\f}$. 
Considering stationarity and the zero angle of attack, the flow is homogeneous in both the azimuthal and temporal directions, the perturbation ansatz can be written,
\begin{align}
    {\fluctuation{\q}} \left(x, r, \theta, t\right)  = \hat{\q}\left( x, r \right) e^{i \left(-\w t + \m \theta \right)} + \pmb{\cc} , \nonumber \\
    ~{\fluctuation{\f}} \left(x, r, \theta, t\right)  = \hat{\f}\left( x, r \right) e^{i\left(-\w t + \m \theta \right)} + \pmb{\cc}, \label{eq:ansatz}
\end{align}
where $\m$ and $\w = 2\pi f$ are, respectively, the azimuthal wavenumber and temporal pulsation, and  $\pmb{\cc}$ denotes the complex conjugate. In this paper, subscript $r$ will refer to the real part of the complex value. Injecting the ansatz \ref{eq:ansatz} in equation~\ref{eq:Theory:reduced_fluct_NS_system} and simplifying, the linear system can be expressed in the frequency domain as:
\begin{align}
    \hat{\q} = \pmb{\R} \hat{\f},
\end{align}

\noindent where $\pmb{\R} =\pmb{\R} \left( \w, \m \right) = \left( i\w \mathsfbi{I} - \mathsfbi{J} \right)^{-1}$ is the resolvent operator. The system defined by equation \ref{eq:Theory:reduced_fluct_NS_system} is thus rewritten as an input/output problem where for every forcing $\hat{\f}$, at a given azimuthal wavenumber and pulsation, there is a corresponding response $\hat{\q}$. The ratio between forcing and response energies defines a gain:
\begin{align}
    \mu^2 \left( \w, \m \right) = \frac{||\hat{\q}||_{q}}{||\hat{\f}||_{f}} 
\end{align}
\noindent where $||\cdot||_{x}$ is the norm associated with a weighted inner product defined as $\langle\pmb q,\pmb q\rangle_{x} = \pmb q^H \mathsfbi{W}_x \pmb q$, where $\mathsfbi{W}_x$ is a weight matrix. To evaluate the perturbation energy, the Chu norm \citep{chu_energy_1965} is generally used for compressible flows. This energy can be written in non-dimensional form following \cite{bugeat_3d_2019}. 
In what follows, both the forcing and response norms use the respective matrix form $\mathsfbi{W}_f$ and $\mathsfbi{W}_q$ of the Chu norm. 
Optimal forcing and response can then be found by maximising the gain,
\begin{align}
    \mu^2_0 \left( \w, \m \right) = \max_{\hat{\f}}\frac{||\hat{\q}||_{{q}}^2}{||\hat{\f}||_{f}^2} = \max_{\hat{\f}} \frac{\hat{\f}^H \pmb{\R}^H \mathsfbi{W}_q \pmb{\R} \hat{\f} }{\hat{\f}^H \mathsfbi{W}_f \hat{\f} }.
\end{align}
The gain defines a generalised Rayleigh quotient which is linked to the generalised hermitian eigenvalue problem:
\begin{align}
    \pmb{\R}^H\mathsfbi{W}_{q}\pmb{\R}\hat{\f} = \mu^2\mathsfbi{W}_{f}\hat{\f}.
\end{align}
Solving this system is equivalent to solving for the right singular vectors of the singular value decomposition (SVD) of the weighted resolvent operator \citep{towne_spectral_2018},
\begin{align}
   \pmb{\R}_w &= \mathsfbi{W}_q^{1/2} \pmb{\R} \mathsfbi{W}_f^{-1/2}  = \mathsfbi{\tilde{Q}} \pmb{\Sigma}\mathsfbi{\tilde{F}}^{H},\\
   \mathsfbi{Q} &= \mathsfbi{W}_q^{-1/2} \mathsfbi{\tilde{Q}}, \\
   \mathsfbi{F} &= \mathsfbi{W}_f^{-1/2} \mathsfbi{\tilde{F}}.
\end{align}
By the properties of the SVD, $\pmb{Q}$ and $\pmb{F}$ form, respectively, orthonormal bases for the optimal responses ${\hat{\q}}_i$ and optimal forcings ${\hat{\f}}_i$ ordered from the most energetic mode to the least, this ordering being manifest in the gains contained in the diagonal matrix $\pmb{\Sigma}$. 
Any real forcing $\hat{\f}_{real}$ experienced in flight or in a wind tunnel would then be expressed as a linear combination of the optimal orthogonal forcing basis and the flow response would be a linear combination of the optimal orthogonal response basis modulated in amplitude by the gains and the inner product $\langle \hat{\f}_i, \hat{\f}_{real}\rangle$, 
\begin{align}
    \hat{\q}_{real} = \sum_{i=0}^{N} \hat{\q}_i \mu_i^2 \langle \hat{\f}_i, \hat{\f}_{real}\rangle, \label{eq:receptivity}
\end{align}

\noindent where $\hat{\q}_i$ and $\hat{\f}_i$ are respectively the $i^{th}$ vector of the optimal response and forcing basis and $\mu_i^2$ the associated gains. 
This leads to a receptivity coefficient $c_i =\mu_i^2 \langle \hat{\f}_i, \hat{\f}_{real}\rangle$ expressing how well a given external disturbance is able to bring energy to the flow response. 
It should be noted that in order to keep the gain values comparable between different cases in the non-dimensionalised framework, the gains can be scaled by the reference timescale.

Contrary to classical linear stability methods, such as LST or PSE, no assumption is made about the baseflow parallelism nor about the modal nature of the instability mechanisms in the streamwise direction.
Resolvent modes account for the non-normality of the linear operator by definition \citep{schmid_nonmodal_2007}. 
This allows for the exploration and ranking of linear amplification mechanisms in the flow without constraints whether they are of modal or non-modal nature. 
If a modal instability exists, such as the second Mack mode, then the resolvent optimal forcing will be the structure that optimally receives energy from external sources (Equation \ref{eq:receptivity}) in order to amplify the instability, and the non-normality of the operator may be leveraged to excite the instability. 
The response will be the spatial structure of this mechanism, the associated gain will be very large due to the exponential modal amplification and the gain separation between leading and sub-optimal modes will be pronounced.

\label{section:Numerical_Strategy}
\subsection{Numerical solver}

Both the baseflow and resolvent analyses are carried out using the BROADCAST solver developed at ONERA \citep{poulain_broadcast_2023}. The code employs a finite-volume formulation, with high-order FE-MUSCL state reconstruction at cell interfaces for the convective fluxes. In this study, a $7^{\text{th}}$-order reconstruction is used. Viscous fluxes are computed using a $4^{\text{th}}$-order compact finite difference scheme. Temporal integration of the Navier–Stokes equations begins from a uniform initial condition and is advanced using an implicit LU-SGS scheme with an approximate Jacobian until transients are eliminated. Final convergence of the steady-state baseflow is achieved through a pseudo-Newton iteration, with a convergence criterion of $\left\|\partial \q / \partial t\right\|_{L_2} \leq 10^{-10}$.

After obtaining the steady baseflow, the linear fluxes are obtained via algorithmic differentiation of the non-linear fluxes using the \emph{source transformation} approach, implemented with the TAPENADE library from INRIA \citep{hascoet_tapenade_2013}. This technique produces linearized fluxes that are exact and retain the formal accuracy of the original non-linear discretization. 
These linearized expressions are used to assemble the Jacobian matrix at the fixed point. Owing to the use of explicit spatial discretization schemes, the resulting Jacobian is both sparse and accurate, making it well suited for the global resolvent analysis.

\subsubsection{Boundary conditions}

BROADCAST uses ghost cells to impose all the boundary conditions on the flow. Supersonic inflow is imposed with a Dirichlet condition in the far field. Axisymmetric conditions are imposed on the axis of axisymmetry. 
Outflow is imposed with a $0^{th}$ order extrapolation at the supersonic outlet. 
Finally, isothermal or adiabatic wall viscous boundary conditions are imposed on all surfaces.

\subsection{Meshing of the numerical domain}
The domain of interest is the $(O, x, r)$ half plane of the blunt cone with an axisymmetric condition. The grid construction procedure follows the previous convergence studies done in \cite{caillaud_separation_2025}.
The mesh is structured and curvilinear and is constructed from mesh-refinement laws imposed on the wall, the shock line, and the far field line in the streamwise direction. 
In the wall normal direction, the mesh-refinement laws  are imposed on the axisymmetry line, the sphere cone junction, and the outlet.
A transfinite interpolation is applied to construct the complete mesh.
The height of the first cell at the wall is set to $0.25$ times the value of a $y^+$ criteria based on the laminar skin friction. About $120$ cells are present in the boundary layer. 
The imposed shock line on the mesh is a spline found in an iterative manner until the shock line is sufficiently well aligned with the physical shock such that numerical convergence of the fixed point ensues.

\section{Baseflow}
\label{section:Baseflow_charateristics}

\begin{figure}
\centering
\begin{subfigure}[t]{0.615\textwidth}
    \captionsetup{width=\linewidth}
    \includegraphics[]{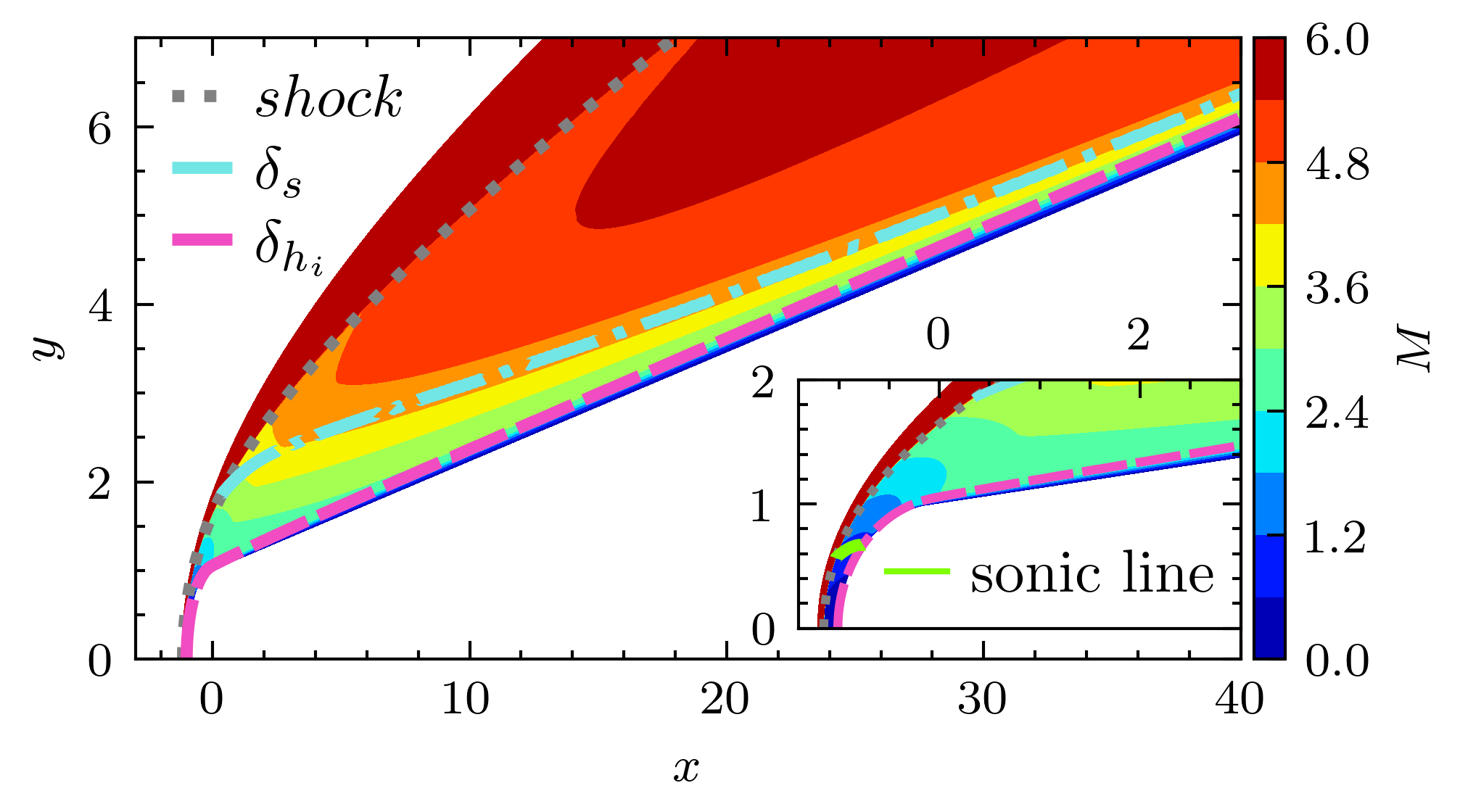}
    \caption{Isothermal blunt cone baseflow}
    \label{subfig:Resultats/BF_small_domain}
\end{subfigure}
\begin{subfigure}[t]{0.19\textwidth}
    \captionsetup{width=\linewidth}
    \includegraphics[]{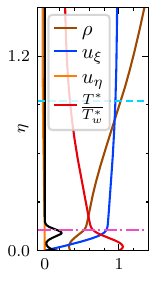}
    \caption{$\xi = 10$ baseflow wall-normal profile}
    \label{subfig:Resultats/BF_slice_xi_Rn10}
\end{subfigure}
\begin{subfigure}[t]{0.15\textwidth}
    \captionsetup{width=\linewidth}
    \includegraphics[]{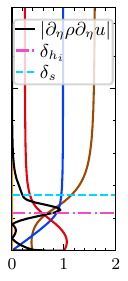}
    \caption{$\xi = 85$ profile}
    \label{subfig:Resultats/BF_slice_xi_Rn85}
\end{subfigure}
\begin{subfigure}[t]{0.35\textwidth}
    \centering
    \includegraphics[valign=t]{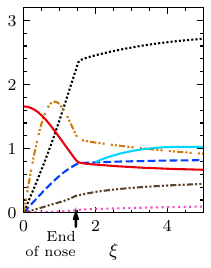}
    \captionsetup{width=\linewidth}
    \caption[]{Boundary layer edge quantities, zoom on the nose section}
    \label{fig:Resultats/BL_quantities_zoom_nose}
\end{subfigure}
\begin{subfigure}[t]{0.60\textwidth}
    \includegraphics[valign=t]{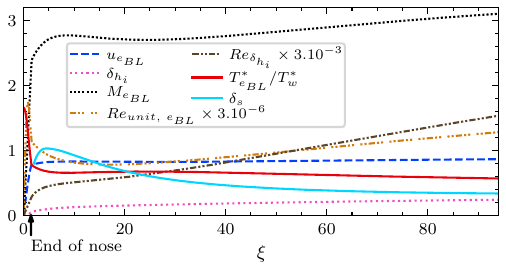}
    \captionsetup{width=\linewidth}
    \caption[]{Boundary layer edge quantities over the full domain}
    \label{fig:Resultats/BL_quantities}
\end{subfigure}
\caption[]{}
\end{figure}

The blunt geometry creates a detached, curved bow shock in front of the object. This generates a continuous range of Rankine-Hugoniot jump conditions where the entropy varies, thus creating an entropy layer. 
Figure \ref{subfig:Resultats/BF_small_domain} displays the Mach number over the isothermal blunt cone where the detached shock is visible along with the entropy layer and boundary layer heights.
Velocity also varies in the entropy layer, as can be seen in Figures \ref{subfig:Resultats/BF_slice_xi_Rn10}, \ref{subfig:Resultats/BF_slice_xi_Rn85}, rendering any definition of the boundary and entropy layer heights delicate. 
To remain consistent with the literature, definitions from \cite{paredes_nose-tip_2019} are used where the boundary layer height ($\delta_{h_i}$) is determined by the minimum height to the wall where the total enthalpy reaches $99.5 \%$  of its freestream value: $h_i \left(\delta_{h_i}\right) = 0.995 h_{i, \infty}$. 
Similarly, the entropy layer height ($\delta_s$) is defined as the minimum height to the wall for which the entropy difference to the freestream is equal to $25\%$ of its value at the wall : $\Delta s \left( \delta_s \right) = 0.25 \Delta s_{wall} $, where $\Delta s \left(x\right) = s \left(x\right)-s_\infty$.

The conditions along the upper limit of the boundary layer are plotted in Figures \ref{fig:Resultats/BL_quantities_zoom_nose} \& \ref{fig:Resultats/BL_quantities}. 
On the nose section, the evolutions of the quantities is fast. The Mach number starts at zero at the stagnation point and increases to about 2.2 at the end of the sphere, crossing the sonic line at $\xi = 0.72$. 
Due to the strong favourable pressure gradient, the boundary layer remains thin over the nose and begins to grow further downstream.
It reaches a maximum at the end of the domain of $\delta_{h_i} = 0.24$.
With the chosen definition for entropy layer height, the upper limit is only reached after $\xi = 2$.
The edge of the entropy layer reaches a maximum at $\xi = 4.5$ with $\delta_s = 1.03$ before decreasing and asymptotically converging towards the boundary layer height. 
\section{Isothermal resolvent gain maps}
\label{section:isothermal_resolvent_gain_maps}

\begin{figure}
\centering
\begin{subfigure}[t]{0.49\textwidth}
    \captionsetup{width=\linewidth}
    \includegraphics[]{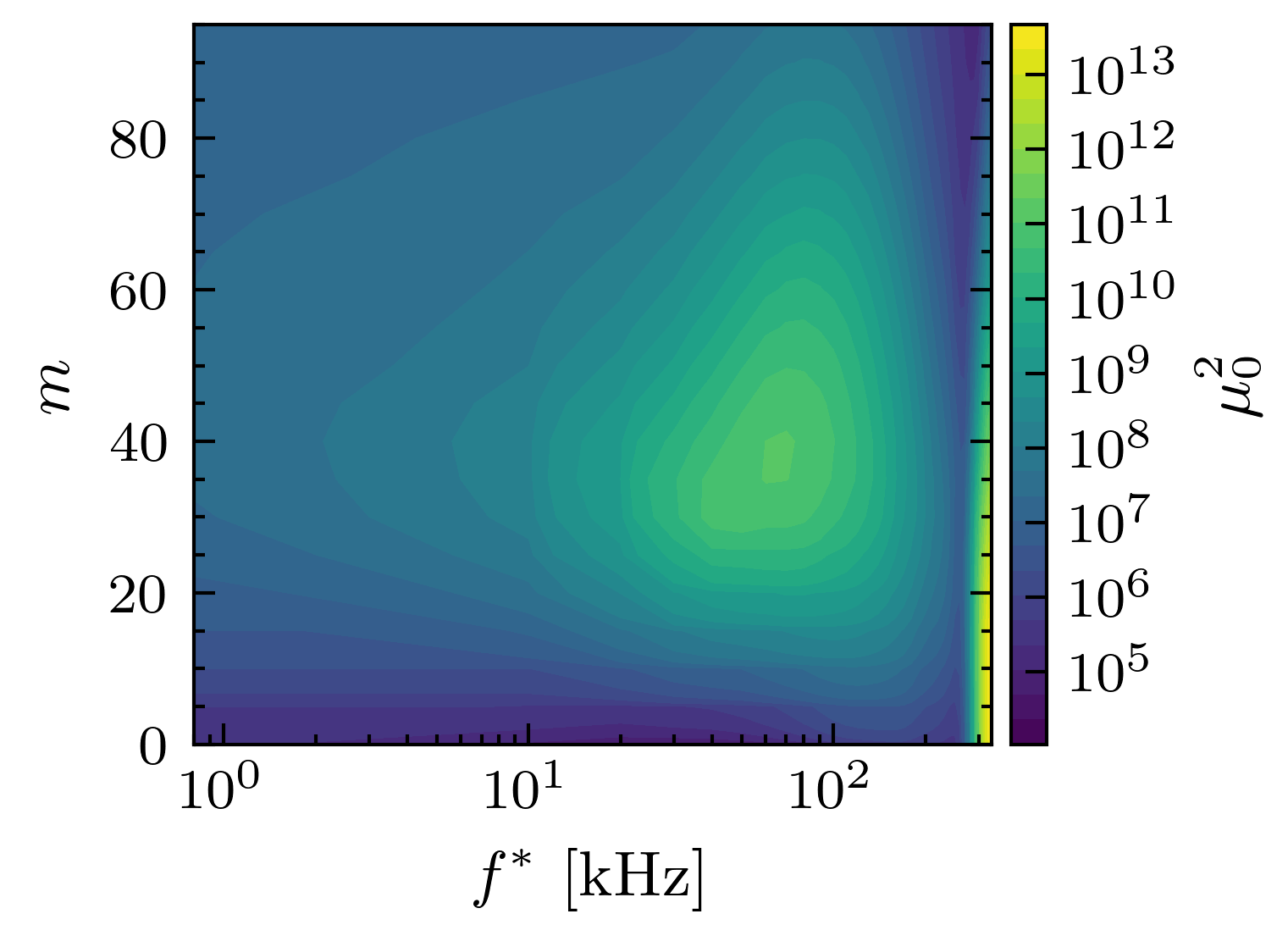}
    \caption{Isothermal sharp cone}
    \label{subfig:Resultats/carte_gain_opt_sharp_iso}
\end{subfigure}
\begin{subfigure}[t]{0.49\textwidth} \includegraphics[]{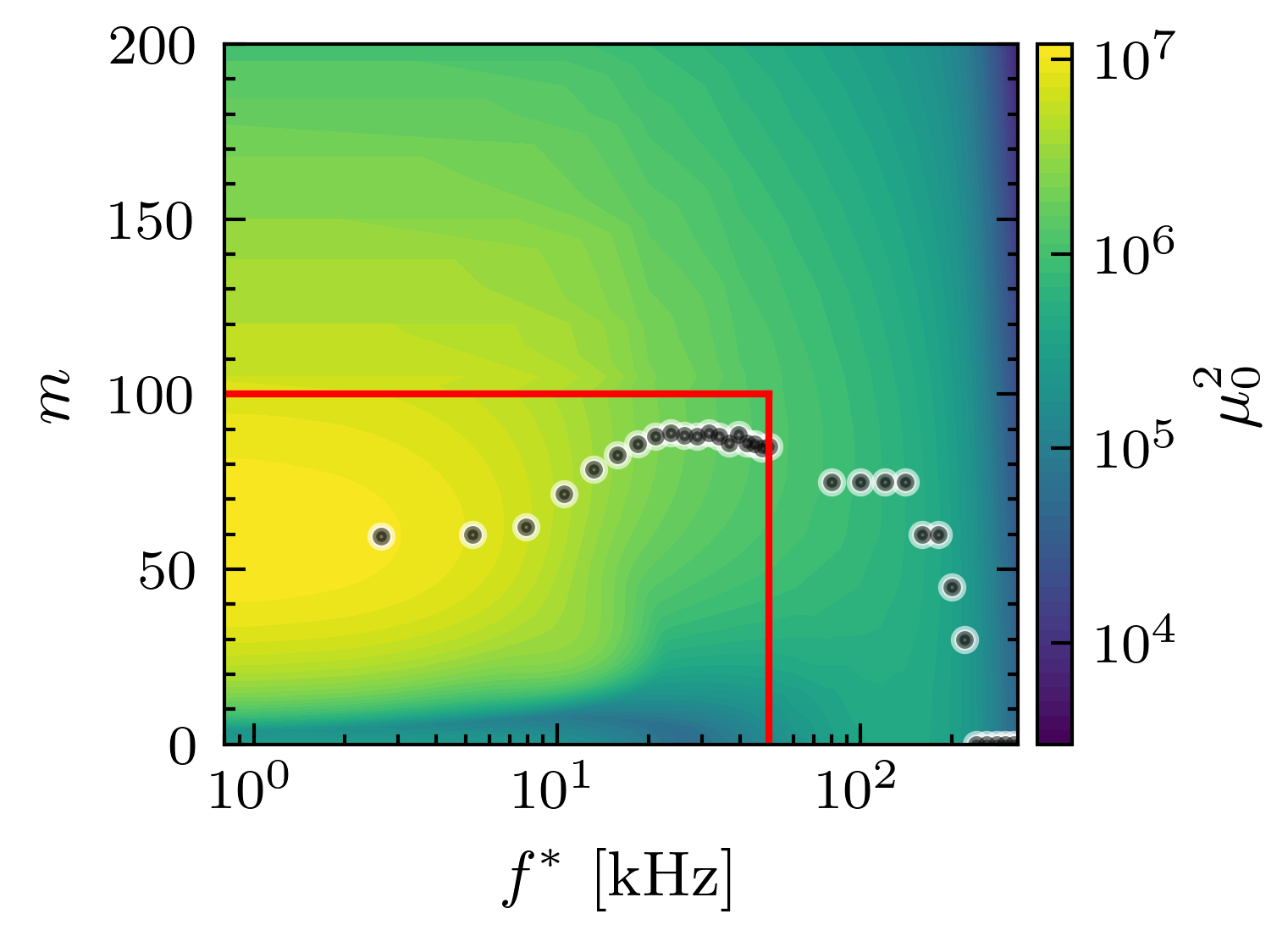}
    \caption{Isothermal blunt cone}
    \label{subfig:Resultats/carte_gain_opt_large_iso_Rn5}
\end{subfigure}
\begin{subfigure}[t]{0.49\textwidth} \includegraphics[]{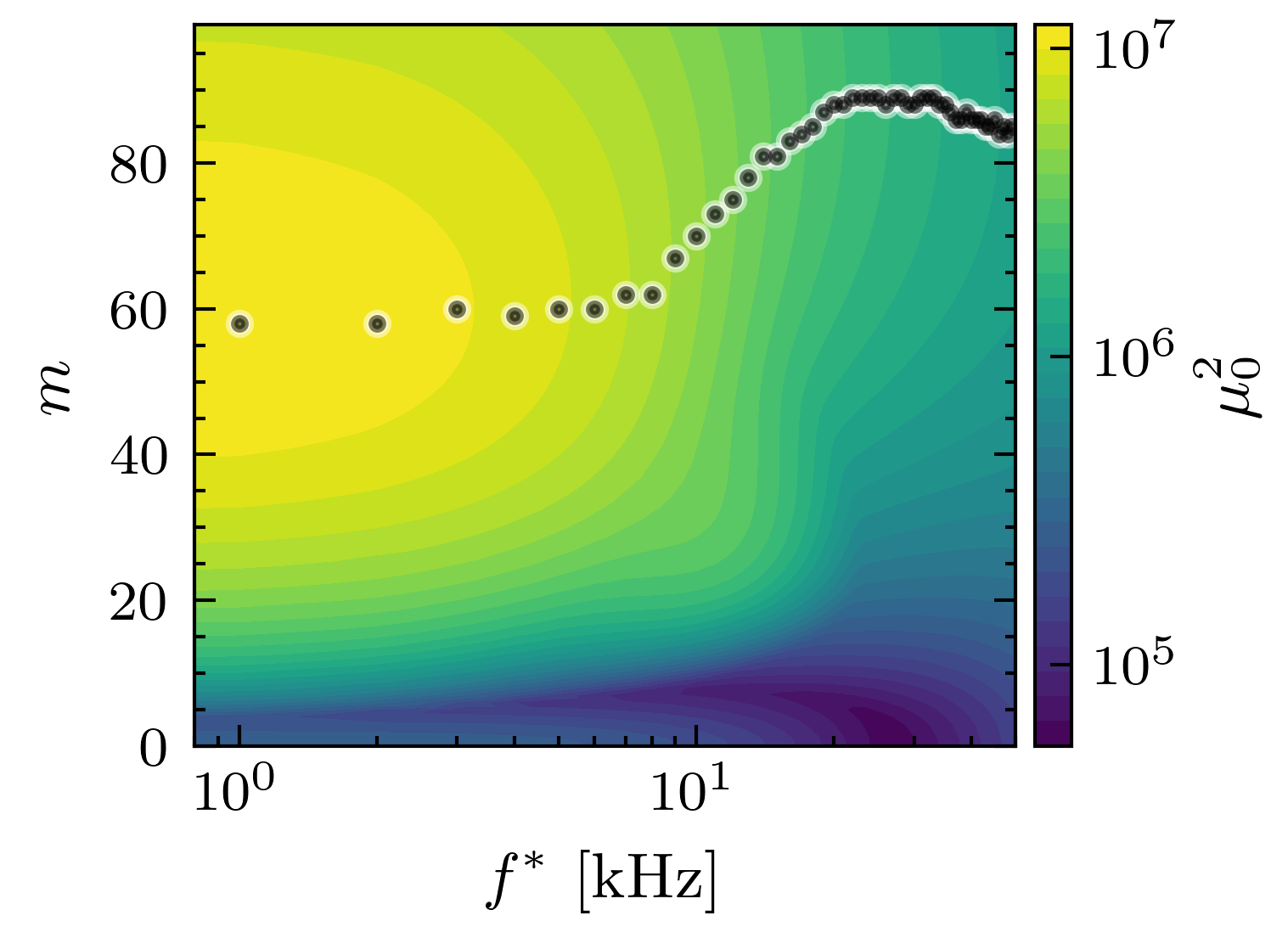}
    \caption{Isothermal blunt cone zoom}
    \label{subfig:Resultats/carte_gain_opt_zoom_iso_Rn5}
\end{subfigure}
\begin{subfigure}[t]{0.49\textwidth}
	\centering \includegraphics[]{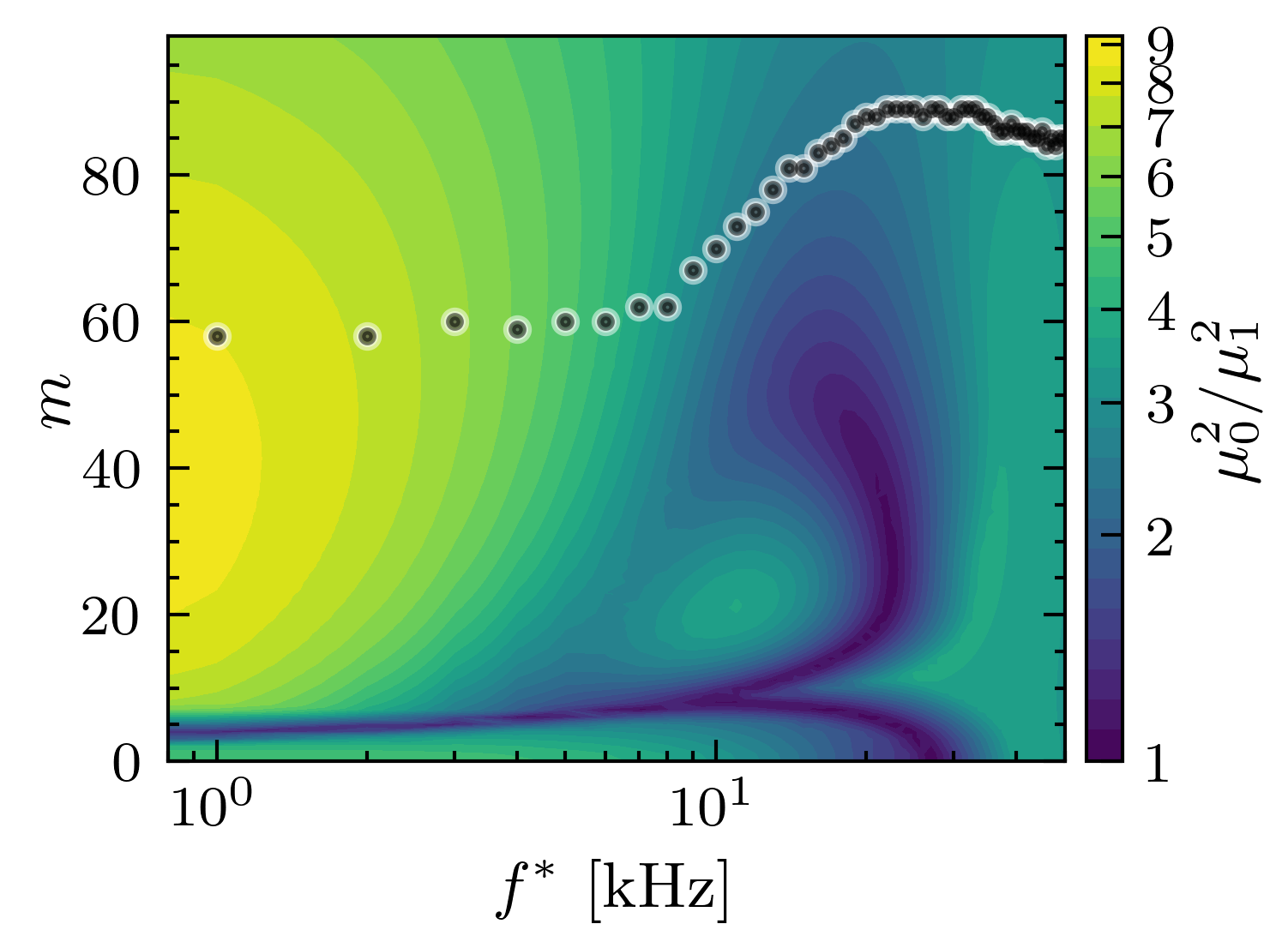}
    \caption{Isothermal blunt cone gain separation map}
    \label{subfig:Resultats/carte_gain_subopt_zoom_iso_Rn5}
\end{subfigure}
\caption{Isothermal gain maps in the $(f^*, m)$ plane. 
\blackdotwwhite ~ most amplified mode at given $f^*$}
\label{fig:Resultats/carte_gain_iso_Rn5}
\end{figure}

A first way to obtain an overview of the dominant amplification mechanisms is to look at optimal gain maps $\mu_0^2(f^*,m)$.
This section presents results for the isothermal blunt-cone case and uses a classical sharp cone as a reference to assess the effect of bluntness. 
Figures \ref{subfig:Resultats/carte_gain_opt_sharp_iso}, \ref{subfig:Resultats/carte_gain_opt_large_iso_Rn5} show the optimal gain map $\mu_0^2$ in the frequency-azimuthal wavenumber plane $(f^*, m)$ for frequencies between 0 and \SI{330}{\kHz} and azimuthal wavenumber ranging from 0 to 100 in the sharp reference and up to 200 for the blunt case. 
The resolvent analysis reveals a stark difference between sharp and blunt cases. 
Firstly, the maximum gain for the blunt case is more than 6 orders of magnitude smaller than the sharp case. 
This large difference can be attributed to the already known strong attenuation of the modal Mack modes, and more specifically the $2^{nd}$ Mack mode, due to bluntness. In the sharp case, this mode is most amplified for $(f^*, m) = (330 ~\SI{}{\kHz}, 0)$ and corresponds to the most amplified mode in the computed $(f^*,m)$ plane. The attenuation in the blunt case is due to the lower local Reynolds number caused by the entropy layer \citep{lei_linear_2012} which weakens the density gradient subduing the acoustic resonance which the second mode uses to amplify \cite{batista_mechanism_2020},
and is consistent with experiments performed by \cite{marineau_mach_2014}. 
In this analysis, the $2^{nd}$ mode is so strongly attenuated in the blunt case that it no longer exists as an optimal resolvent mode (see Appendix \ref{appA} for more details about mesh convergence). 
The dynamics shift from being dominated be a lower-rank high-frequency mechanisms to a predominantly higher-rank system where the most amplified modes are at lower frequencies. Figure \ref{subfig:Resultats/carte_gain_opt_zoom_iso_Rn5} shows a zoom between $f^* \in ]0, 50] ~\kHz$ and $m \in [0, 100]$ of the gain map corresponding to the red rectangle in Figure \ref{subfig:Resultats/carte_gain_opt_large_iso_Rn5}. Figure \ref{subfig:Resultats/carte_gain_subopt_zoom_iso_Rn5} presents the corresponding gain separation between the optimal mode and the first suboptimal mode, expressed as the ratio $\mu_0^2/\mu_1^2$. 

\begin{figure}
\centering
\begin{subfigure}[t]{0.49\linewidth}
    \includegraphics[width = 1.1\linewidth]{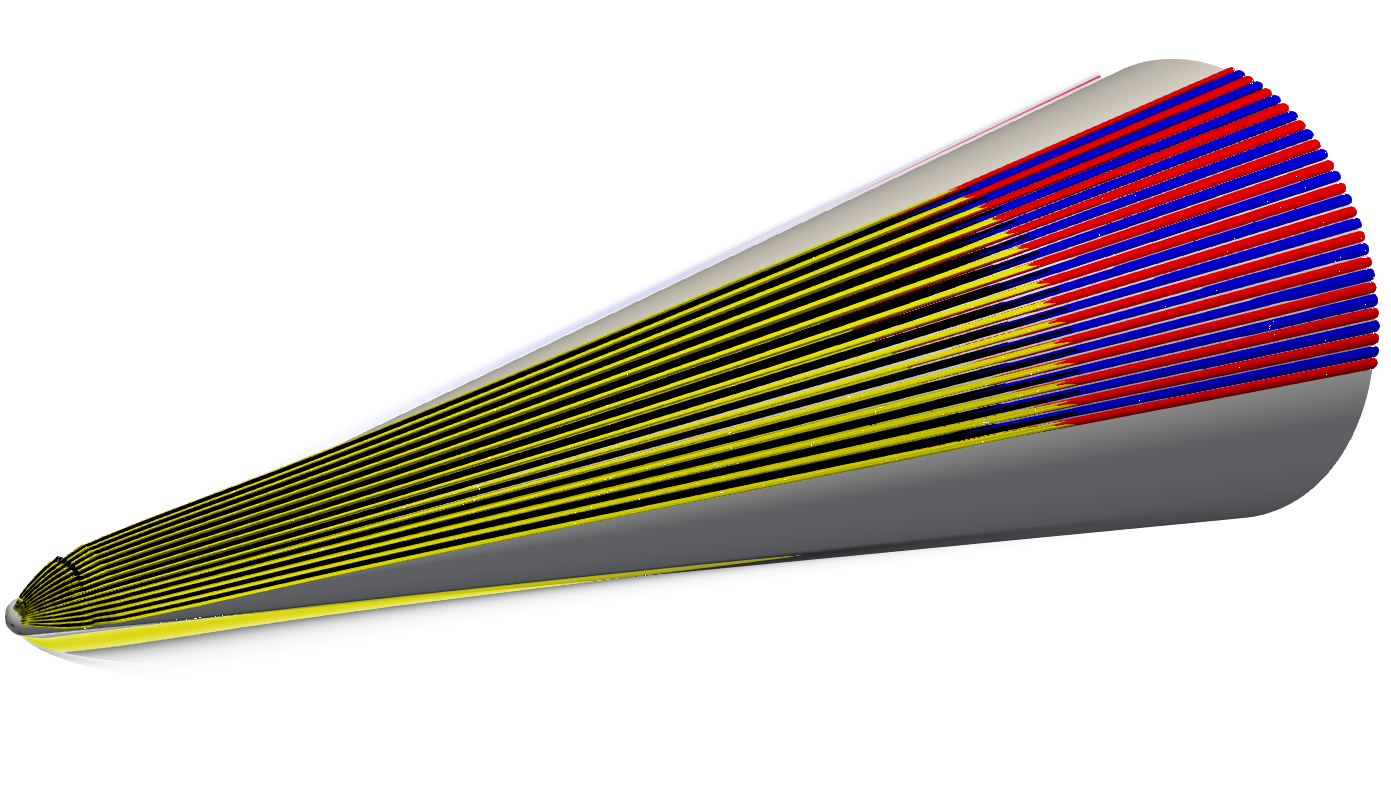}
    \captionsetup{width =  0.9\linewidth}
    \caption{3D visualisation of the optimal steady streak, $(f^*, m) = (0, 57)$, in black \& yellow an isolevel of the forcing and in red \& blue an isolevel of the response} 
    \label{fig:Resultats/3D_f0_m57_streak_iso_mode_no_bg}
\end{subfigure}
\begin{subfigure}[t]{0.49\linewidth}
    \centering
     \hspace*{-0.1\linewidth}\includegraphics[width = 1.1\linewidth]{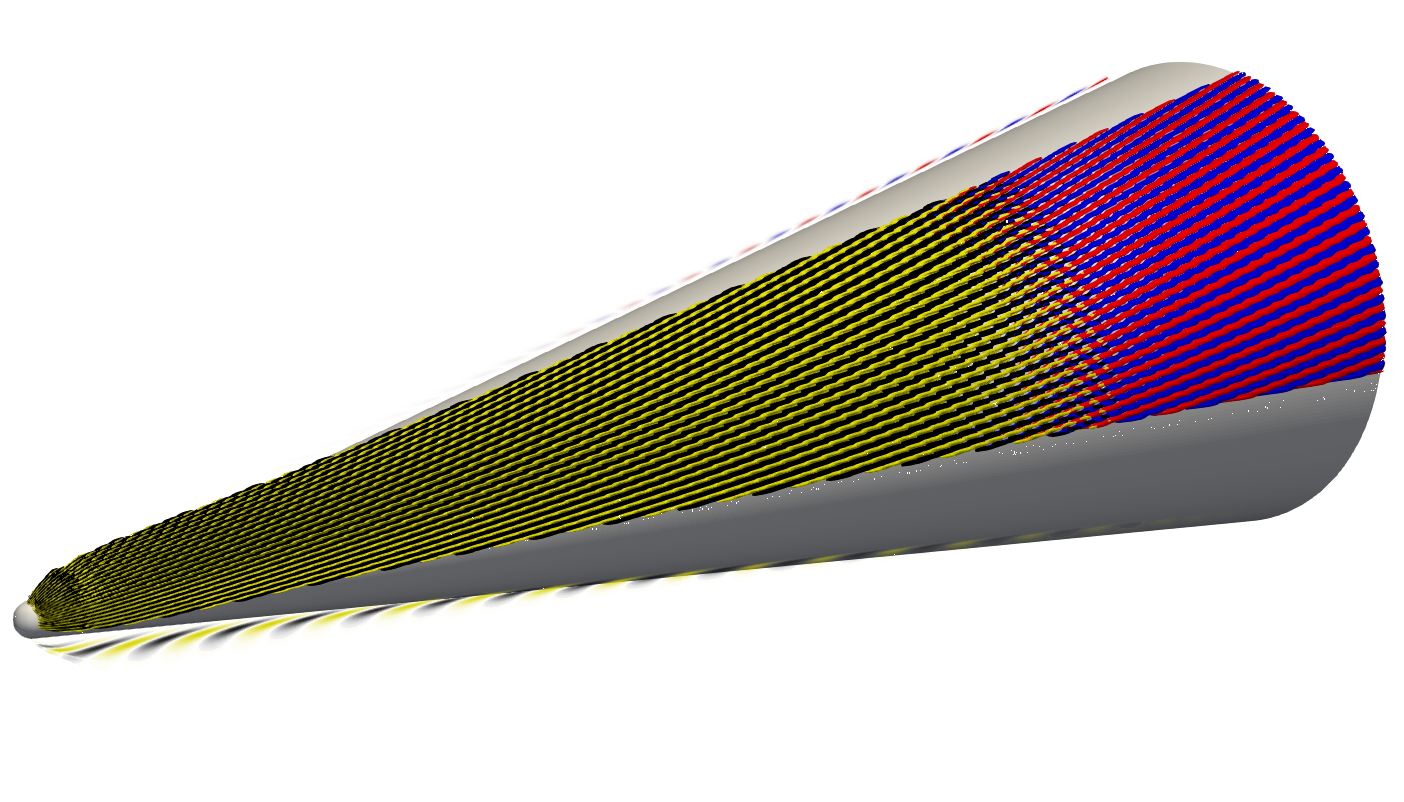}
    \captionsetup{width =  0.9\linewidth}
    \caption{3D visualisation of the $(f^*, m) = (35 ~\kHz, 88)$ HAWEL mode, in black \& yellow an isolevel of the forcing and in red \& blue an isolevel of the response} 
    \label{fig:Resultats/3D_f35_m88_entropy_layer_mode_no_bg}
\end{subfigure}
\caption{}
\end{figure}

The family of most amplified modes correspond to streaks at $f^* = 0~ \SI{}{\kHz}$ with the most amplified at $m = 57$ (Figure \ref{fig:Resultats/3D_f0_m57_streak_iso_mode_no_bg}) with gain ${\mu_0^2} \approx 10^{7}$ \SI{}{\square\s}, and gain separation around $9$. 
This is consistent with the PSE analysis of \cite{paredes_nonmodal_2019}, which finds streaks to be the most amplified mode in a configuration close to ours.
This mode will be discussed in detail later, in section \ref{section:OptimalStreakMode}.

\begin{figure}
\centering
\includegraphics[width = 0.42\linewidth]{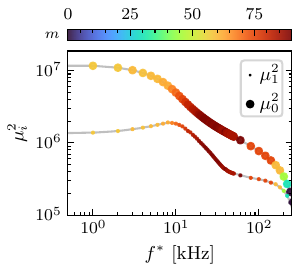}
\captionsetup{}
\caption[]{Optimal and $1^{st}$ suboptimal gain functions for the most amplified modes as function of frequency for the isothermal blunt case}
\label{fig:Resultats/most_amplified_modes}
\end{figure}

The most amplified mode as a function of frequency, given by the black dots, shifts towards higher azimuthal wavenumber as frequency increases and reaches a maximum of $m = 89$ at $f^* = 21 ~\SI{}{\kHz}$. 
The azimuthal wavenumber of the most amplified mode remains high up to $f^* = 140 ~\SI{}{\kHz}$ before dropping to zero. 
These modes between \SI{20}{\kHz} and \SI{140}{\kHz} have a strong forcing and response support in the entropy layer and will be referred to as high-azimuthal-wavenumber-entropy-layer modes (HAWEL) (Figure \ref{fig:Resultats/3D_f35_m88_entropy_layer_mode_no_bg}). 
These results further agree with \cite{paredes_nonmodal_2019} who observed a strong amplification of entropy layer modes. 
The gain separation for this family of modes ranges between 2 and 4 as can be seen in Figure \ref{subfig:Resultats/carte_gain_subopt_zoom_iso_Rn5}.
The gain of the most amplified mode decreases with frequency, as shown in Figure \ref{fig:Resultats/most_amplified_modes}. 
For the frequency range corresponding to HAWEL modes, the gain is $6$ to $20$ times less than the optimal streak mode. 
However, entropy layer wisps have been observed experimentally \citep{grossir_influence_2019, kennedy_characterization_2022, ceruzzi_experimental_2024}, motivating a  thorough investigation of their properties , which we present in Section \ref{section:entropy_layer_mode_iso}.

In the gain separation map, Figure \ref{subfig:Resultats/carte_gain_subopt_zoom_iso_Rn5}, a local peak appears at $(f^*, m) = (11 ~\SI{}{\kHz}, 21)$. The optimal mode at this position corresponds to the $1^{st}$ Mack mode, which can be recognised by the characteristic forcing and response profiles (see Figure \ref{fig:Resultats/sub_slice_m_21}, in Appendix \ref{Appendix:suboptimal_profiles}). 
As mentioned, due to the change in baseflow induced by bluntness, the first Mack modes are significantly damped and shifted to lower frequencies and azimuthal wavenumbers compared to the sharp cone case, where the most amplified first Mack mode occurs at $(f^*, m) = (65 ~\SI{}{\kHz}, 37)$.

\subsection{Suboptimal modes}
\label{section:suboptimal}

The higher-order nature of the blunt cone flow regime, (i.e. lower gain separation), in comparison with the low-order modal mechanisms associated with the sharp cone, motivates a more detailed review of the suboptimal resolvent response and forcing modes. In this section, the gain separation is discussed for different azimuthal wavenumbers and the associated response and forcing structures are investigated.

\begin{figure}
\centering
\begin{subfigure}[t]{0.32\textwidth} \centering \includegraphics[]{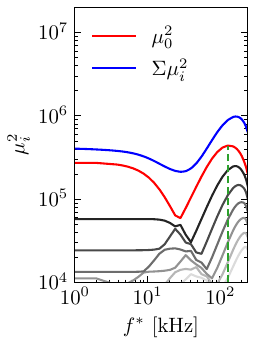}
\captionsetup{}
    \caption{$m = 0$}
    \label{subfig:Resultats/gains_and_subgains_vs_f_m0}
\end{subfigure}
\begin{subfigure}[t]{0.32\textwidth} \centering \includegraphics[]{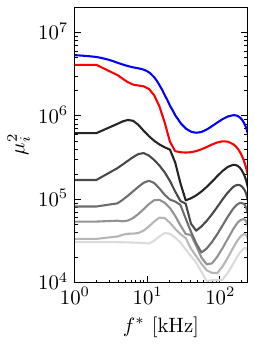}
\captionsetup{}
    \caption{$m = 21$}
    \label{subfig:Resultats/gains_and_subgains_vs_f_m21}
\end{subfigure}
\begin{subfigure}[t]{0.32\textwidth} \centering \includegraphics[]{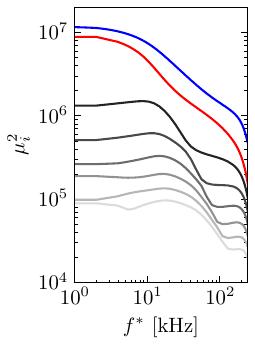}
\captionsetup{}
\caption[]{$m = 88$}
\label{subfig:Resultats/gains_and_subgains_vs_f_m88}
\end{subfigure}
\caption{Optimal gain and first few suboptimal gains as function of $f$ for different $m$}
\label{fig:Resultats/gains_and_subgains_vs_f_all_m}
\end{figure}

\begin{figure}
\centering
\begin{subfigure}[t]{\textwidth} 
\centering
\includegraphics[]{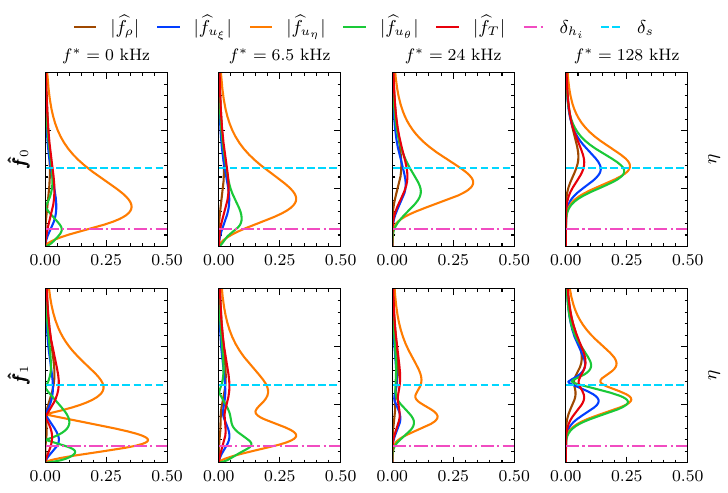}
\caption[]{Absolute value of optimal forcing (top, mode 0) and $1^{st}$ suboptimal forcing (bottom, mode 1) at $m = 88$ at $\xi = 20$} 
\label{subfig:Resultats/sub_slice_m_forcing_88}
\end{subfigure}
\begin{subfigure}[t]{\textwidth}
\centering
\includegraphics[]{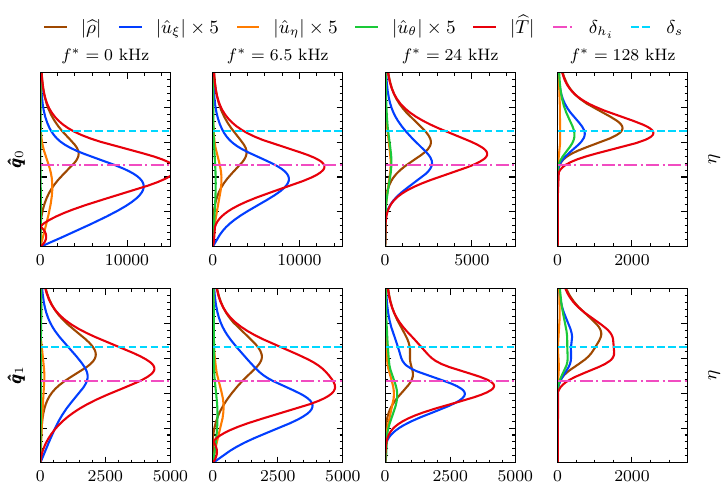}
\caption[]{Absolute value of optimal response (top, mode 0) and $1^{st}$ suboptimal response (bottom, mode 1) at $m = 88$ at $\xi = 94$} 
\label{subfig:Resultats/sub_slice_m_response_88}
\end{subfigure}
\caption[]{ }
\label{fig:Resultats/sub_slice_m_88}
\end{figure}

In Figure \ref{fig:Resultats/gains_and_subgains_vs_f_all_m} is presented the gain functions of the first few optimal modes at different azimuthal wavenumbers.
Starting at $m = 88$, the gains of the first two optimal modes are presented in Figure \ref{subfig:Resultats/gains_and_subgains_vs_f_m88}. 
Figure \ref{fig:Resultats/sub_slice_m_88} shows the profiles of the first two optimal modes at different frequencies at $\xi = 20$ for the forcing and at the end of the domain for the response.
At low frequency, the most amplified mode is a suboptimal streak (the optimal being at $m = 57$). 
This is observable with the characteristic shape of the forcing structure with wall-normal and azimuthal velocity forcing and a strong response on streamwise velocity and temperature. 
The key characteristics for the optimal streaks will be discussed in more detail in section \ref{section:OptimalStreakMode}.

As the frequency increases (Figure \ref{subfig:Resultats/gains_and_subgains_vs_f_m88}), the gain diminishes and the forcing structure moves into the entropy layer for both the optimal and suboptimal modes : these are the HAWEL modes.
At $f^* = 128$ \SI{}{\kHz}, the forcing signs most strongly on $f_{\hat{\Vel}_\theta}$ and $f_{\hat{\Vel}_\xi}$ and the response responds strongly on temperature in the entropy layer.
The difference between the optimal and suboptimal mode is the presence of a two lobe structure in the forcing, however the response profiles are quite similar.
A more detailed analysis of HAWEL modes will be done in section \ref{section:entropy_layer_mode_iso}.

\begin{figure}
\centering
\begin{subfigure}[t]{\textwidth} 
\centering
\includegraphics[]{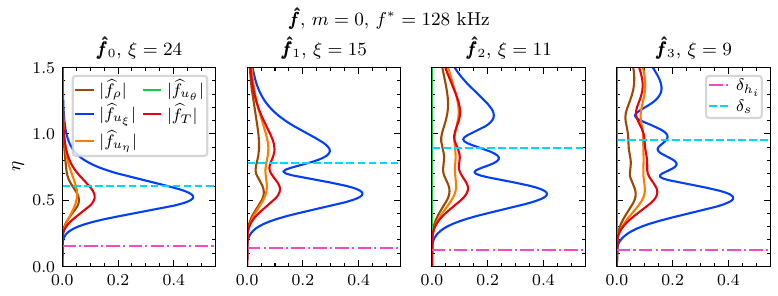}
\captionsetup{width=\linewidth}
\caption[]{Absolute value of optimal forcing profiles of the planar modes at $m = 0$, $f^* = 128\SI{}{\kHz}$ of the 4 first optimal modes at the curvilinear coordinate of maximum streamwise velocity forcing} 
\label{subfig:Resultats/sub3_128kHz_slice_forcing_m0_ximax_U}
\end{subfigure}
\begin{subfigure}[t]{\textwidth}
\centering
\includegraphics[]{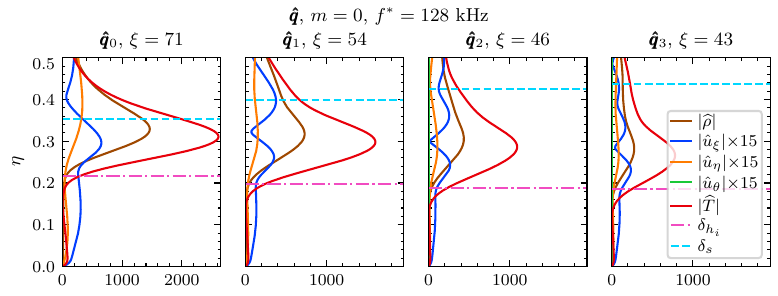}
\captionsetup{width=\linewidth}
\caption[]{Absolute value of optimal response profiles of the planar modes at $m = 0$, $f^* = 128\SI{}{\kHz}$ of the 4 first optimal modes at the curvilinear coordinate of maximum temperature fluctuation} 
\label{subfig:Resultats/sub3_128kHz_slice_response_m0_ximax_T}
\end{subfigure}
\caption[]{ }
\label{fig:Resultats/sub3_128kHz_slic}
\end{figure}

Now considering the modes at $m = 0$, see Figure \ref{subfig:Resultats/gains_and_subgains_vs_f_m0}, there appear to be two clear mechanisms switching dominant position at $f^* = 25$ \SI{}{\kHz}. 
The gain for the lower frequency region ($f^*\leq 10$\SI{}{\kHz}) correspond to a weak mechanism that has a signature at the shock, suggesting some spurious modes (see Appendix \ref{Appendix:suboptimal_profiles}). 
At higher frequencies, the dominant mechanism switches and the new optimal mode peaks at $f^* = 128$ \SI{}{\kHz} and signs in the entropy layer.
Figure \ref{fig:Resultats/sub3_128kHz_slic} shows the wall-normal profiles of the first 4 optimal modes at $128$ \SI{}{\kHz}, following the green dashed line in Figure \ref{subfig:Resultats/gains_and_subgains_vs_f_m0}.
The streamwise position of these modes is the position at which there is maximum amplitude in the streamwise velocity forcing and temperature response which are the dominant primitive variables of these modes.
All these modes have a forcing and response signature in the entropy layer.
As the mode order increases, both the peak forcing and response position move upstream. 
Though these modes sign in the entropy layer they have characteristics different from HAWEL modes that distinguishes them : the most intense forcing variable is on streamwise velocity forcing contrary to HAWEL modes where it is a combination of wall-normal and azimuthal velocity forcing.

At $ m = 21$, see Figure \ref{subfig:Resultats/gains_and_subgains_vs_f_m21}, there is again a shift in optimal mechanism. 
At low frequency the optimal mode is a streak. 
When the frequency increases the optimal mode becomes the $1^{st}$ Mack mode and has the largest gain separation around $f^* = 11$ \SI{}{\kHz}. 
At $f^* = 20$ \SI{}{\kHz} there is a switch in dominant amplification mechanism. 
These new modes have maximum gain near $f^* = 130 ~\kHz$ and 
have similar characteristics compared to the $(f^* = 128 ~\kHz, m = 0)$ modes but have increased azimuthal velocity forcing.
These modes between \SI{25}{\kHz} and \SI{240}{\kHz} and at low azimuthal wavenumber seem to be oblique version of the planar modes earlier discussed and will be referred to as Low Azimuthal Wavenumber Entropy Layer modes (LAWEL), see Appendix \ref{Appendix:suboptimal_profiles} for more details.

A summary of these mechanisms main features is provided in Table~\ref{tab:modes-features}. The next sections will focus on the description and analysis of the mechanisms associated with the growth of the dominant optimal modes, starting with the optimal streak mode.

\begin{table}
    \centering
    \begin{tabular}{lllll}
         & Streaks & 1st-mode & LAWEL & HAWEL\\\hline
        Frequency range [\SI{}{\kHz}]& $[0,5]$ & $[7,20]$ & $[70,250]$  & $[20,140]$ \\
        $m$ range &$[40,80]$ & $[15,35]$ & $[0,20]$ & broad $m_{peak}\approx 90$  \\
        Gain $[s^2]$ & $\sim 10^7$ & $\sim 2.10^6$ & $\sim 4.10^5$ & $[6.10^5, 2.10^6]$\\
        Maximum gain separation & $\approx 10$ & $\approx 5$  & $\approx 4$ & $\approx 4$ \\\hline
    \end{tabular}
    \caption{Summary of the main mechanisms}
    \label{tab:modes-features}
\end{table}

\section{Isothermal blunt cone optimal streak mode}
\label{section:OptimalStreakMode}

\begin{figure}[t]
\centering
\begin{subfigure}[t]{0.48\textwidth} \includegraphics[]{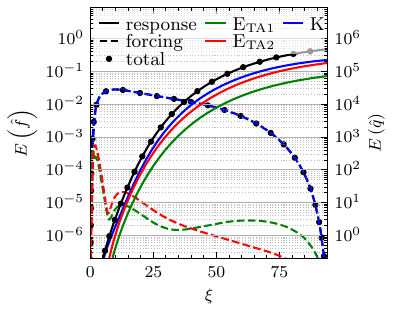}
	\caption{}
    \label{fig:Resultats/E_chu_streaks}
\end{subfigure}
\begin{subfigure}[t]{0.48\textwidth} \includegraphics[]{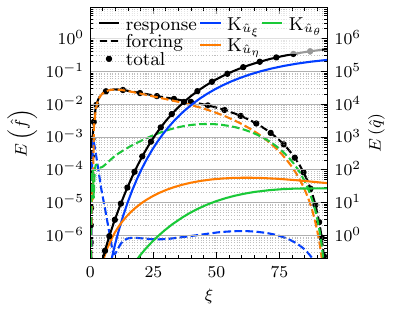}
	\caption{}
	\label{fig:Resultats/E_chu_streaks_momentum}
\end{subfigure}
\caption{Wall-normal integrated Chu energy density ($E_\eta$) (left) and kinetic energy density (right) as function of $\xi$ of the streak mode}
\end{figure}

The overall most amplified mode for a unit forcing energy corresponds to a steady streak mode at $f^* = 0$ \SI{}{\kHz} and $m=57$.
In order to explore the characteristics of this streak mode in comparison with sharp cone flows, an analysis of the Chu energy is conducted. This energy norm is defined as
\begin{align}
\label{eq:chu_energy}
E_{Chu} = \frac{1}{2} \int_V{\underbrace{\mean{\Density} \left| \fluctuation{\Vvec} \right|^2}_{\text{Kinetic (KE)}} +\underbrace{\frac{\mean{\Temp}}{\mean{\Density}\gamma \M^2} {\fluctuation{\Density}}^2}_{\text{Density term (D)}} + \underbrace{\frac{\mean{\Density}}{(\gamma-1)\gamma \M^2 \mean{\Temp} } {\fluctuation{\Temp}}^2}_{\text{Temperature term (T)}}dV},
\end{align}
\noindent which consists of a kinetic energy (KE) term, a density term (D) and a temperature term (T).
The wall-normal integrated Chu energy density ($E_\eta(\xi) = \int_0^{\eta_{max}} E\left(\xi, \eta \right) d \eta$) is shown in Figure \ref{fig:Resultats/E_chu_streaks}.
It shows where and how the Chu energy density is distributed for both the forcing and the response. 
The forcing energy density comprises only kinetic energy, the temperature fluctuation T and density fluctuation D energy terms being negligible.
The wall-normal-integrated energy density of the forcing in the nose section is very small, and it strongly increases at the beginning of the cone, peaking at $\xi = 10$, which is slightly after the maximum separation between the entropy layer and boundary layer limits. 
The response is maximum at the end of the domain where it is comprised of kinetic energy fluctuations associated with streamwise streaks, but with a strong contribution from the T energy term. 
Figure \ref{fig:Resultats/E_chu_streaks_momentum}, shows the contribution of the different velocity terms to the kinetic energy. 
The forcing is initially almost entirely dominated by the wall-normal velocity fluctuations, up to around $\xi = 50$ where the kinetic energy of the azimuthal velocity component becomes comparable. 
This later part is the signature of a forcing structure that has taken the form of longitudinal vorticity; but it is interesting that in the upstream region of maximum forcing, the optimal motion involves a strong radial fluctuation localised in the entropy layer.
For the response, it is the streamwise component of the kinetic energy which is most important.

\begin{figure}
\centering
\includegraphics[width = \linewidth]{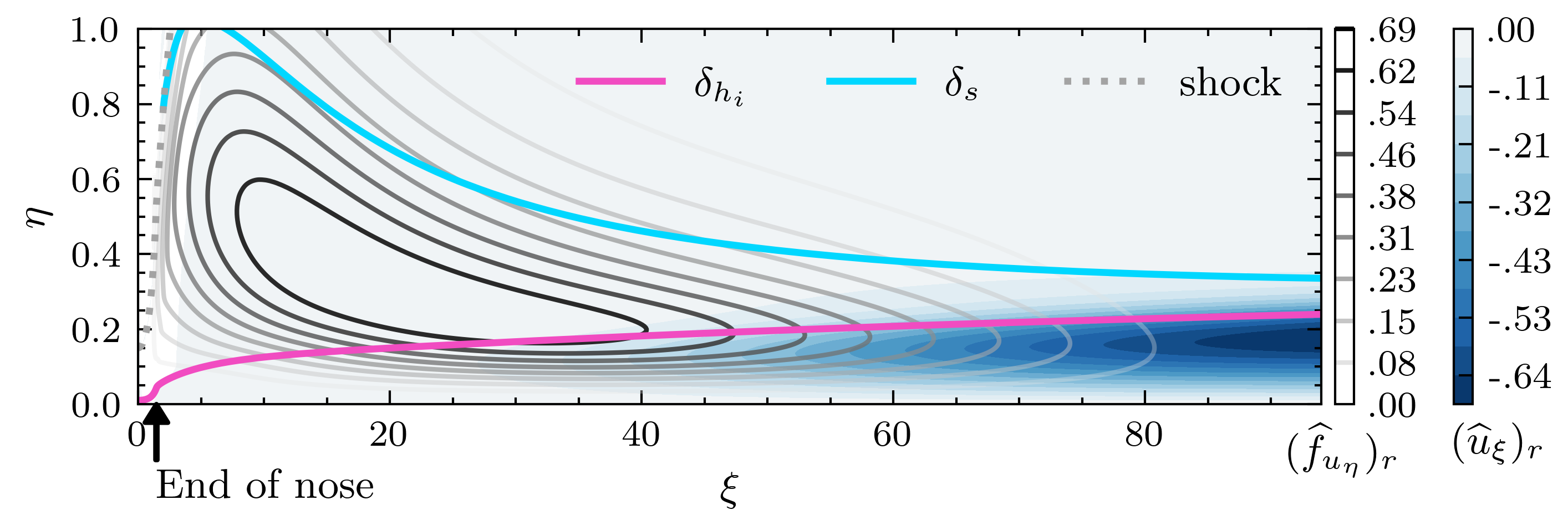}
\caption[]{Global shape of the real parts of the wall-normal velocity forcing (grey) and streamwise velocity response (blue) corresponding to variables with the largest amplitudes in the streak mode. Normalised by their respective maximum absolute value} 
\label{fig:Resultats/domain_streak_Vf_U}
\end{figure}

\begin{figure}
\centering
\begin{subfigure}{0.245\textwidth} 
\captionsetup{width = \linewidth}
\includegraphics[]{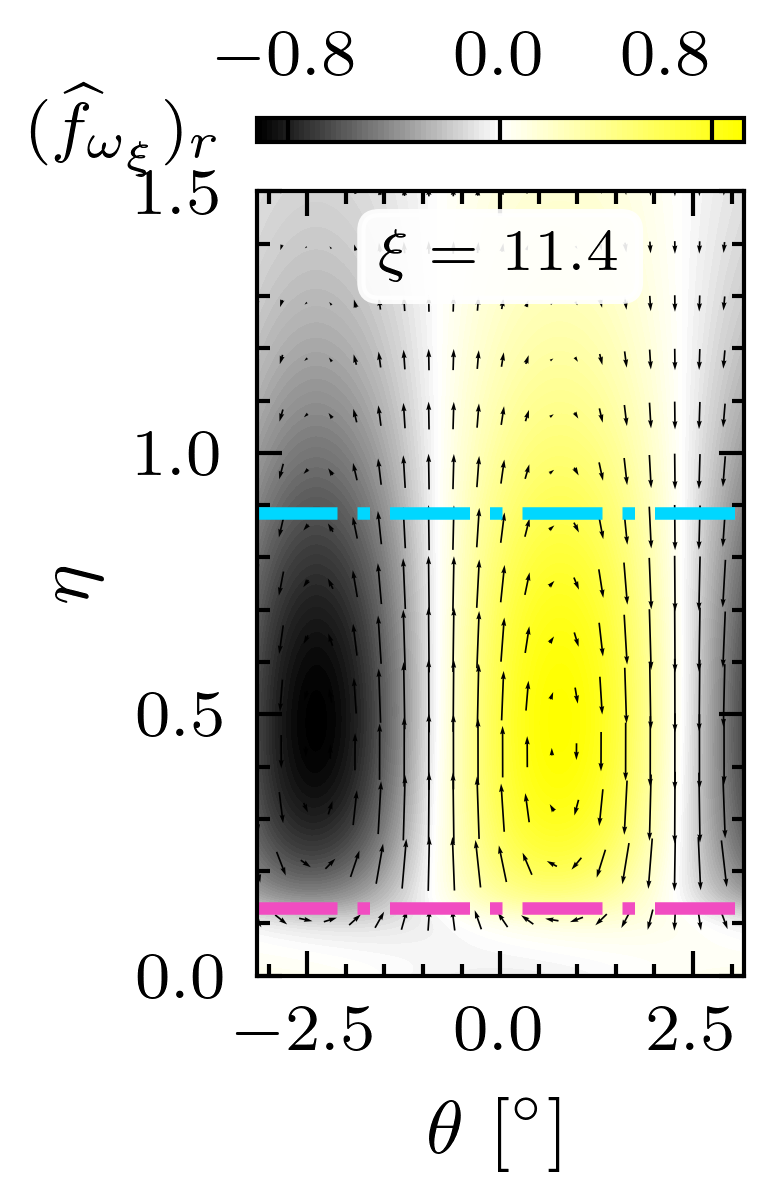}
\caption{Forcing}
\label{subfig:Results/azimuthal_streak_f0_m57_forxing_11}
\end{subfigure}
\begin{subfigure}{0.245\textwidth} 
\captionsetup{width = \linewidth}
\includegraphics[]{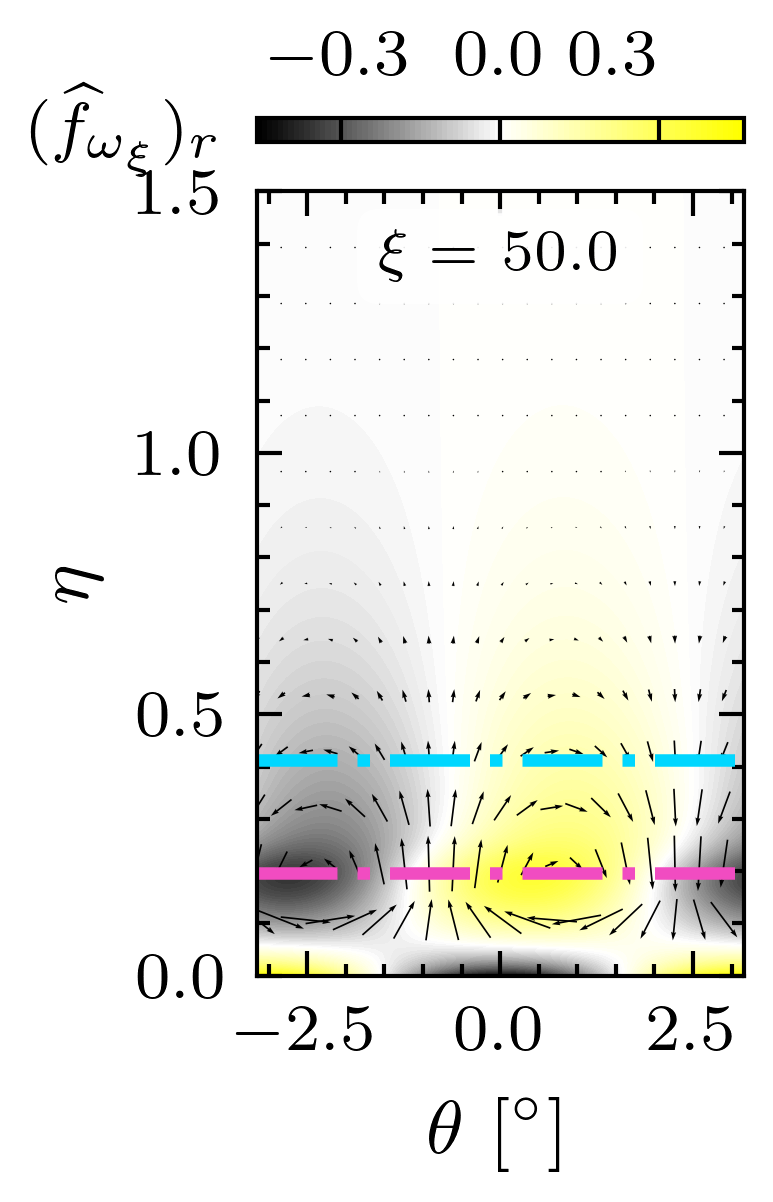}
\caption{Forcing}
\label{subfig:Results/azimuthal_streak_f0_m57_forxing_50}
\end{subfigure}
\begin{subfigure}{0.245\textwidth} 
\captionsetup{width = \linewidth}
\includegraphics[]{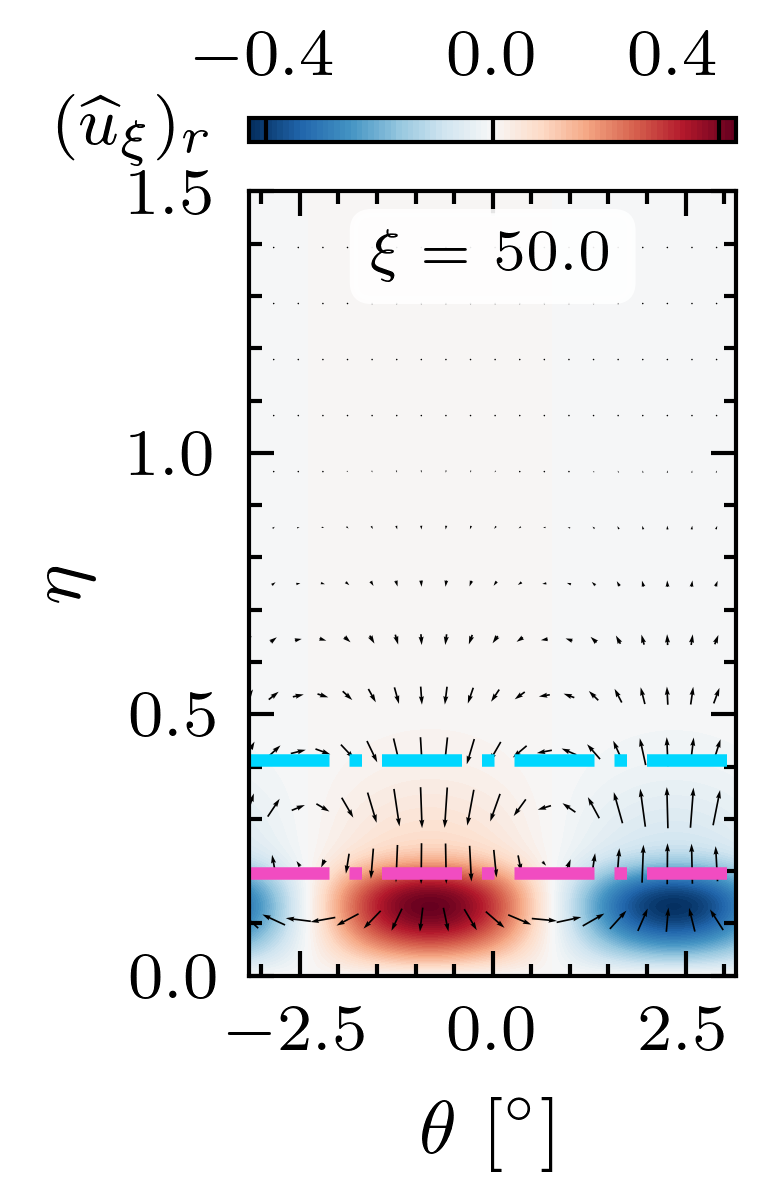}
\caption{Response}
\label{subfig:Results/azimuthal_streak_f0_m57_response_50}
\end{subfigure}
\begin{subfigure}{0.245\textwidth}
\captionsetup{width = \linewidth}
\includegraphics[]{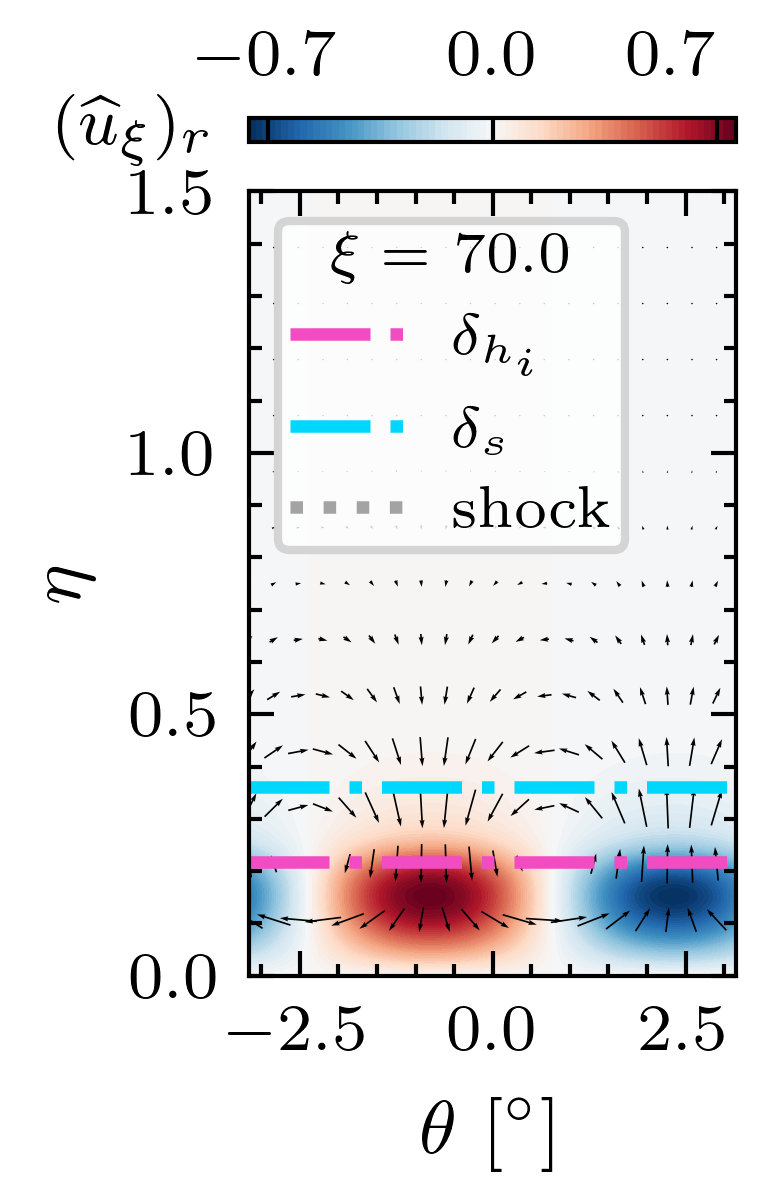}
\caption{Response}
\label{subfig:Results/azimuthal_streak_f0_m57_response_70}
\end{subfigure}
\captionsetup{width= 0.93\linewidth}
\caption[]{Real part of the forcing velocity vector field with associated streamwise vorticity \textbf{(left)} \& real part of the response velocity \textbf{(right)} for the optimal streak mode}
\label{fig:Results/azimuthal_streak_f0_m57_response}
\end{figure}

The forcing and response structures are shown in the $(\xi, \eta)$ streamwise plane in Figure \ref{fig:Resultats/domain_streak_Vf_U} for the most energetic Chu norm components: $\hat{f}_{\Veta}$ and $\hat{\Vel}_\xi$.
Additionally, to get a better picture of how the mode is organised, multiple $\left(\eta, \theta \right)$ cross-sections at different axial positions are shown in Figure \ref{fig:Results/azimuthal_streak_f0_m57_response} for both forcing and response. 
These figures show the forcing velocity field and the associated streamwise vorticity, computed as $\hat{\f}_{\pmb{\omega},\xi} = \partial \hat{\pmb f}_\theta / \partial \eta - \partial \hat{\pmb f}_\eta/\partial \theta$, at two upstream positions where the forcing is strong; and the response velocity vector field further downstream.

Initially, the forcing is dominated by almost purely wall-normal velocity fluctuations which are primarily localised in the entropy layer, and are composed of adjacent azimuthal segments with opposed wall normal velocity.  
At these upstream stations, indeed where receptivity is maximal (peak in forcing), the mechanism has many features comparable with the lift-up mechanism, but a key difference lies in the dominant forcing support being localised in the entropy layer which lies well above the boundary layer in the upstream region. 
This highlights the importance of the entropy layer in the generation of streaks for blunt bodies. 
And it provides an indication that atmospheric disturbances traversing the bow shock can excite streak-like disturbances within the boundary layer without necessarily penetrating as far as the boundary layer.
The downstream evolution of forcing and response illustrate how, as the entropy layer merges with the boundary layer, the characteristic roll-streak configuration of the lift-up mechanism forms. The longitudinal-vorticity of the forcing gradually moves into the boundary layer, reinforcing the streak-like response in the boundary layer, which continues to grow with increasing streamwise position.

To explore the mechanisms responsible for the creation of this mode, we consider the energy budget of the linearised kinetic and temperature energy equations, similarly to \cite{dwivedi_reattachment_2019}. 
In this analysis, the energy (KE) and temperature (T) terms are decomposed into transport (Tr), production (P), source (S), viscous (V) and forcing (F) terms as following:
\begin{align}
    \mean{\Density} \partial_t {\fluctuation{\Vvec}} \cdot \fluctuation{\Vxi}  \mathbf{e}_\xi+  \text{KE}_{\text{Tr}, \xi} &= \text{KE}_{\text{P1}, \xi} + \text{KE}_{\text{P2}, \xi} + \text{KE}_{\text{S}, \xi} + \text{KE}_{\text{F}, \xi} \label{eq:KE_equation}\\
    \frac{\mean{\Density}}{2(\gamma-1)\gamma M_{\infty}^2 \mean{\Temp}}  \partial_t {\fluctuation{\Temp}}^2 + \text{T}_{\text{Tr}} &=  \text{T}_{\text{I}} +  \text{T}_{\text{II}} + \text{T}_{\text{S1}} + \text{T}_{\text{V1}} + \text{T}_{\text{V}\mu 1} \label{eq:TE_equation}\\ 
    & ~+ \text{T}_{\text{S2}} + \text{T}_{\text{V2}} + \text{T}_{\text{V}\mu 2} + \text{T}_{\text{F}} \nonumber
\end{align}
\noindent the complete derivation if these equation is given in Appendix \ref{Appendix:Linearised_eq}. 

\begin{figure}
\centering
\begin{subfigure}[t]{0.32\textwidth} \includegraphics[]{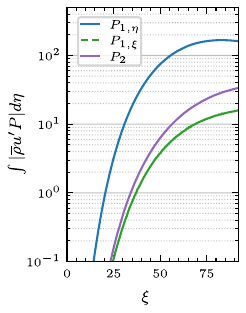}
	\caption{}
	\label{subfig:Resultats/KE_production_mode_f0_m57}
\end{subfigure}
\begin{subfigure}[t]{0.32\textwidth} \includegraphics[]{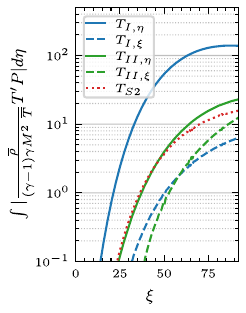}
	\caption{}
	\label{subfig:Resultats/T_production_mode_f0_m57}
\end{subfigure}
\caption{Wall-normal integrated dominant production terms for the linearised kinetic energy equation (left, \ref{subfig:Resultats/KE_production_mode_f0_m57}) and temperature equation (right, \ref{subfig:Resultats/T_production_mode_f0_m57}) for the optimal streak mode}
\label{fig:Resultats/Production_response_f0_m57}
\end{figure}

The streamline curvature is found to be small where the response is located : $1/R < 10^{-10} $, where $R$ is local radius of curvature (Appendix \ref{Appendix:curvature}), this is why the energy budget analysis was conducted on wall-parallel lines.
Figure \ref{fig:Resultats/Production_response_f0_m57} presents the most significant wall-normal integrated, time and azimuthally averaged, energy budget terms as function of $\xi$, here are their expressions for the kinetic energy equation : 
\begin{align}
     \text{KE}_{\text{P1, } \xi} &= \left< - \overline{\Density} \left( \fluctuation{\Vvec}\pmb{\cdot} \nabla \right) \mean{\Vvec}  \pmb{\cdot} \fluctuation{\Vxi} \mathbf{e}_\xi \right> = \left< \overline{\Density}(\underbrace{- \fluctuation{\Vxi} \partial_{\xi} \mean{\Vxi} }_{\text{Acceleration P1}_\xi} \underbrace{-\fluctuation{\Veta} \partial_\eta \mean{\Vxi} }_{\text{Shear P1}_\eta} ) \fluctuation{\Vxi} \right> \label{eq:KE_P1} \\
     \text{KE}_{\text{P2, } \xi} &= \left<- \overline{\Density} \frac{\fluctuation{\Density}}{\mean{\Density}} \left( \mean{\Vvec}\pmb{\cdot} \nabla \right) \mean{\Vvec} \pmb{\cdot} \fluctuation{\Vxi} \mathbf{e}_\xi \right> \approx \left< \overline{\Density}\left( - \frac{\fluctuation{\Density}}{\mean{\Density}} \left( \mean{\Vvec}\pmb{\cdot} \nabla \right) \mean{\Vxi}\right) \fluctuation{\Vxi} \right> \label{eq:KE_P2}
\end{align}

\noindent where $\left< \cdot \right> =  \frac{mf}{2\pi}\int_{0}^{\eta_{\max}}\int_{0}^{2\pi/m}  \int_0^{1/f} \cdot dt d\theta d\eta$, when $m\ne0$ and $f\ne0$, otherwise their respective mean is dropped.

The production terms arise from (i) the transport of baseflow velocity by velocity fluctuations, see Equation \ref{eq:KE_P1}, and (ii) a density fluctuation term, see Equation \ref{eq:KE_P2}. 
Since the kinetic energy of this mode is dominated by streamwise velocity fluctuations, only those are considered in this study. 
The first production term is further separated into two components: the first called \emph{acceleration} is due to the streamwise acceleration or deceleration of the mean flow, and the second called \emph{shear} is dependant on wall-normal gradient of the mean flow and resulting from the lift-up effect, \citep{landahl_note_1980}. 
Figure \ref{subfig:Resultats/KE_production_mode_f0_m57} shows that the lift-up mechanism dominates and is the main mechanism through which kinetic energy is extracted from the baseflow to the mode along the whole domain.

For the T equation, the dominant terms are :
\begingroup
\allowdisplaybreaks
\begin{align}
     \text{T}_{\text{I}} &= \left< - \frac{\mean{\Density}}{(\gamma-1) \gamma M_{\infty}^2 \mean{\Temp}} \fluctuation{\Vvec} \pmb{\cdot} \left( \nabla \mean{\Temp}\right) \right> = \left< - \frac{\mean{\Density}}{(\gamma-1) \gamma M_{\infty}^2 \mean{\Temp}}  ( \underbrace{\fluctuation{\Vxi} \partial_\xi \mean{\Temp}}_{\text{wall-parallel T}_{I,~\xi}} + \underbrace{\fluctuation{\Veta} \partial_\eta \mean{\Temp}}_{\text{wall-normal T}_{I,~\eta}} ) \right>, \label{eq:T_P1} \\
    \text{T}_{\text{II}} &=  \left< - \frac{\fluctuation{\Density}}{(\gamma-1) \gamma M_{\infty}^2 \mean{\Temp}} \mean{\Vvec} \pmb{\cdot} \left( \nabla \mean{\Temp}\right) \right> = \left< - \frac{\fluctuation{\Density}}{(\gamma-1) \gamma M_{\infty}^2 \mean{\Temp}} (\underbrace{ \mean{\Vel}_\xi \partial_\xi \mean{\Temp}}_{\text{wall-parallel T}_{II,~\xi}} + \underbrace{\mean{\Vel}_\eta \partial_\eta \mean{\Temp}}_{\text{wall-normal T}_{II,~\eta}}) \right>, \label{eq:T_P2} \\
    \text{T}_{\text{S2}} &=  \left< - \frac{\mean{\Density}}{ \gamma^2 M_{\infty}^4 } (\nabla \pmb{\cdot} \fluctuation{\Vvec}) \right> , \label{eq:T_S2}
\end{align}
\endgroup
\noindent where $\text{T}_{\text{I}}$ and $\text{T}_{\text{II}}$ represent transport of baseflow temperature by, respectively, velocity and density fluctuations and are the analogous production terms P1 and P2 in the KE equation. 
$\text{T}_{\text{S2}}$ is the velocity dilatation heating.
Here, it is the wall-normal $\text{T}_{\text{I}, \eta}$ term that contributes the most to T energy, see Figure \ref{subfig:Resultats/T_production_mode_f0_m57}. 
The wall-normal velocity response mixes the thermal boundary layer, bringing high baseflow temperature flow up and  low temperature flow down, creating temperature streaks.

In conclusion, the streak receptivity mechanism of the optimal streak modes in the \textit{moderately} blunt regime is comprised of radial velocity pumping in the entropy layer which activates: the lift-up mechanism responsible for the growth of streamwise velocity streaks and a similar mixing mechanism associated with the thermal gradient, creating temperature streaks.
\section{Isothermal blunt cone optimal HAWEL modes}
\label{section:entropy_layer_mode_iso}

While the HAWEL mode family is 6 to 20 times less amplified than the optimal streak mode, entropy layer fluctuations have been observed experimentally by \cite{kennedy_characterization_2022} and controlled DNS studies have shown that they may underpin transition \citep{hartman_nonlinear_2021, paredes_mechanism_2020}. 
Furthermore, \emph{realisable-perturbation} resolvent analysis shows that the gain values computed by standard resolvent, as we are doing, should be considered with caution \citep{kamal_global_2023}. In other words, the mechanisms that will arise depend to a large degree on the kinds of perturbations entering the flow from the atmosphere and how they can project on the optimal forcing support (see Equation~\ref{eq:receptivity}). This is another reason why it is important, in our study, to carefully consider all modes and mechanisms.
For these reasons, a thorough characterisation of HAWEL modes is performed in this section, first on a specific example at $(f^*, m) = (35~ \SI{}{\kHz}, 88)$ and then for the family of HAWEL modes with support in the entropy layer. 
In the following section, we will refer to the HAWEL modes only as entropy layer modes.

\subsection{Analysis of an entropy layer mode at \texorpdfstring{$f^* =$}~ 35 kHz, m = 88}
\label{subsection:entropy_layer_f35_m88}
\begin{figure}[t]
\centering
\begin{subfigure}[t]{0.48\textwidth} \includegraphics[]{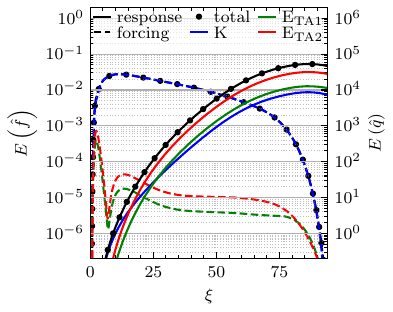}
    \caption{}
    \label{subfig:Resultats/E_chu_entropy_layer_mode}
\end{subfigure}
\begin{subfigure}[t]{0.48\textwidth} \includegraphics[]{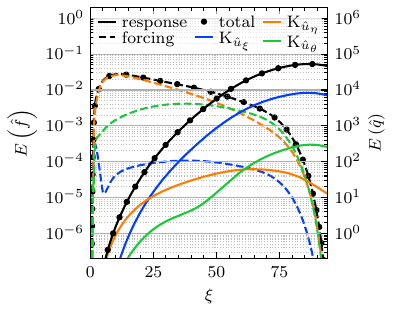}
    \caption{}
    \label{subfig:Resultats/E_chu_entropy_layer_momentum}
\end{subfigure}
\caption{Wall-normal integrated Chu energy density $E_\eta$ (left \ref{subfig:Resultats/E_chu_entropy_layer_mode}) and kinetic energy density (right \ref{subfig:Resultats/E_chu_entropy_layer_momentum}) as function of $\xi$ for the entropy layer mode}
\end{figure}

In this section, a detailed analysis of an entropy layer mode at $f^*$ = 35 kHz, $m$ = 88 is presented. This mode is representative of the larger group of optimal entropy layer modes. And this mode in particular was chosen because it lies well within the entropy layer family mode (i.e far from other optimal mechanisms seen in the gain map Figure \ref{subfig:Resultats/carte_gain_opt_zoom_iso_Rn5}) while being one of the most amplified entropy layer modes with a gain $\mu_0^2 = 1.7\times 10^6$ \SI{}{\square\s}.
The nature of this mode is investigated through its energetics by first looking at its energy growth in the domain. 
Then the shape of this mode is presented.
Finally, an energy budget analysis is carried in order to understand the sources of this energy growth. 

The forcing energy $E_\eta$ is predominantly comprised of kinetic energy, see Figure \ref{subfig:Resultats/E_chu_entropy_layer_mode}. 
There is little forcing energy at the tip of the blunt nose, but it rises sharply at the beginning of the cone section and peaks at $\xi = 11.6 $, close to the maximum separation between the boundary and entropy layers. 
The forcing energy then slowly decreases in intensity, up to $\xi = 70$ where it falls sharply. 
In Figure \ref{subfig:Resultats/E_chu_entropy_layer_momentum}, the kinetic energy density is decomposed into its three velocity components. 
The forcing is primarily composed of wall-normal velocity, until $\xi = 50 $ where azimuthal velocity forcing becomes as important. 
This is a signature of longitudinal vorticity forcing.
The response on the other hand is initially dominated by kinetic energy, up to $\xi = 15$, which is also where the local forcing is most important, but then quickly switches to being dominated by contributions from the temperature fluctuation T energy term in the Chu norm, until the end of the cone. 
At $\xi = 25$, the density D energy term also becomes more important than kinetic energy. 
This suggests that the forcing first excites velocity fluctuations in the entropy layer, and that there is a gradual transfer to density and temperature fluctuations.
The response then peaks at $\xi = 86.4$ before the end of the domain, indicating that the mode is no longer able to grow once the entropy layer reaches some threshold of radial separation or parallelism with the boundary layer.

\begin{figure}[t]
    \begin{subfigure}{\textwidth}
        \includegraphics[width = \linewidth]{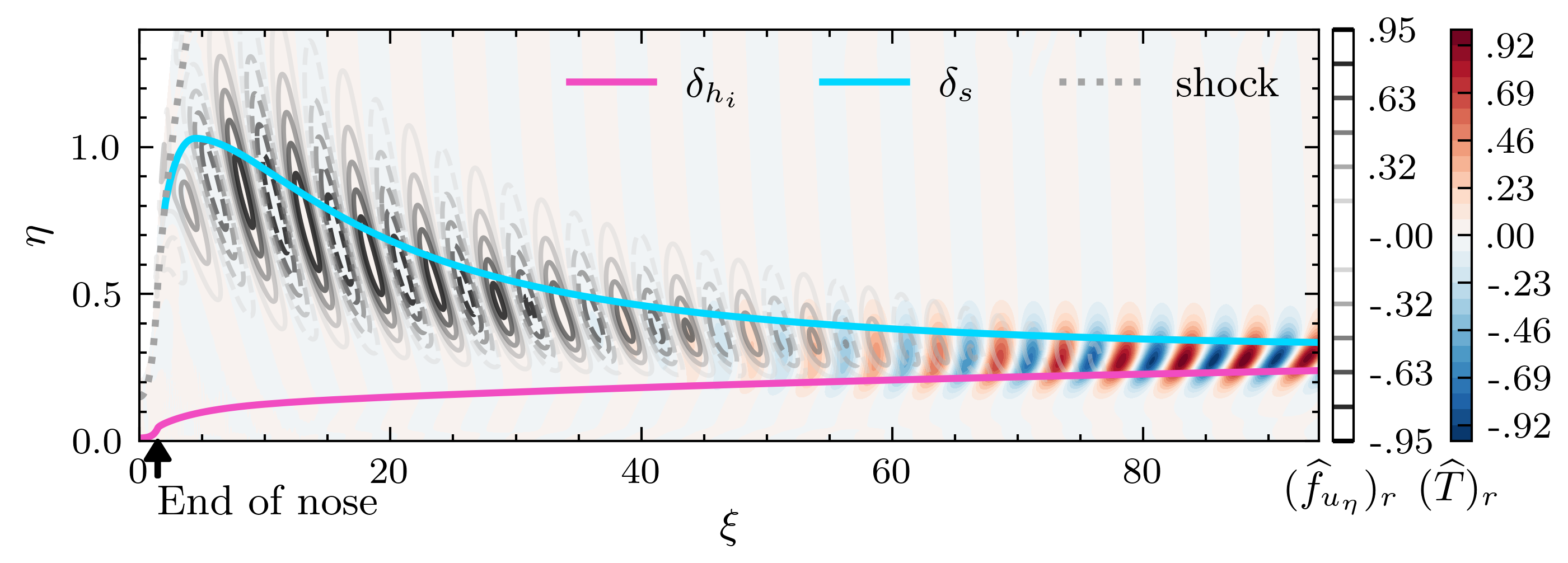}
        \caption[]{Wall-normal velocity forcing}
        \label{fig:Resultats/domain_entropy_layer_f35_m88_Vf_T}
    \end{subfigure}
    \\
    \begin{subfigure}{\textwidth}

    \includegraphics[width = \linewidth]{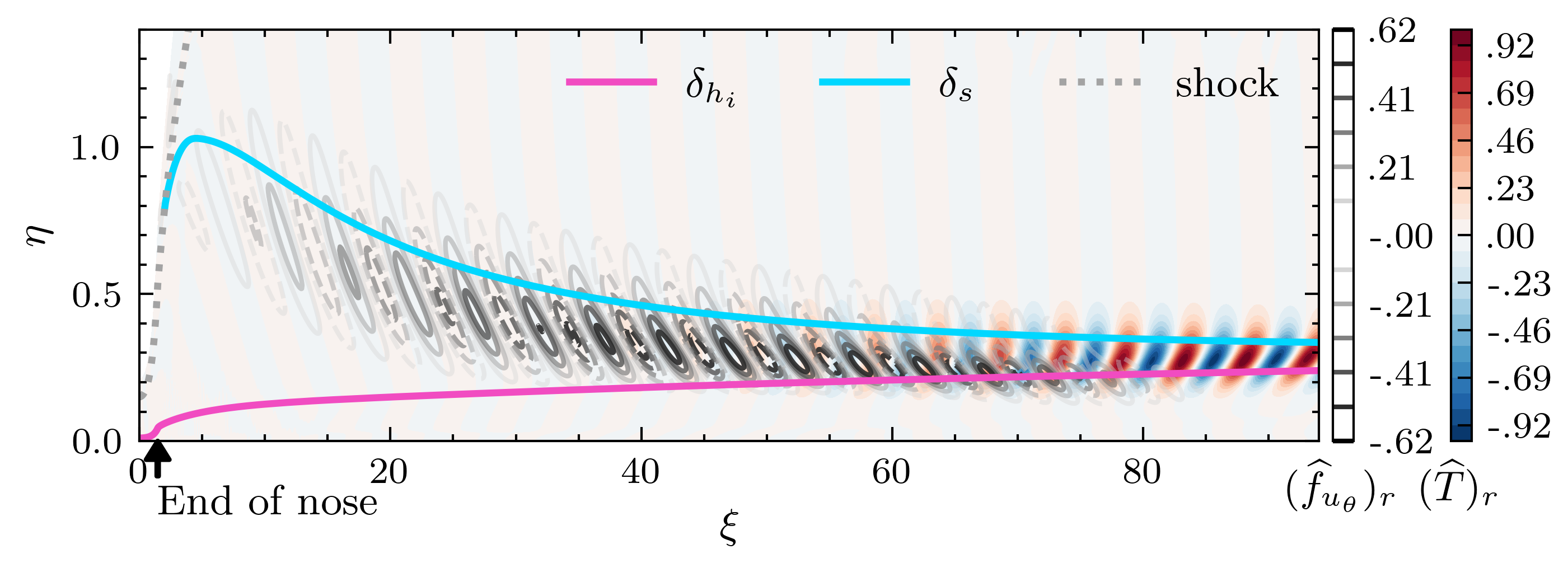}
    \caption[]{Azimuthal velocity forcing}
    \label{fig:Resultats/domain_entropy_layer_f35_m88_Wf_T}
    \end{subfigure}
    \caption[]{Global shape of the real parts of the azimuthal and wall-normal velocity forcing (grey lines, dashed lines are negative) and temperature response (blue) corresponding to variables with the largest amplitudes entropy layer mode. Normalised by $\max ||\hat{\f}_{\Vel}||$ and $\max |\hat{\Temp}|$ respectively}
\label{fig:Resultats/domain_entropy_layer_f35_m88_all}
\end{figure}

\begin{figure}[t]
\centering
\begin{subfigure}{0.245\textwidth} 
\captionsetup{width = \linewidth}
\includegraphics[]{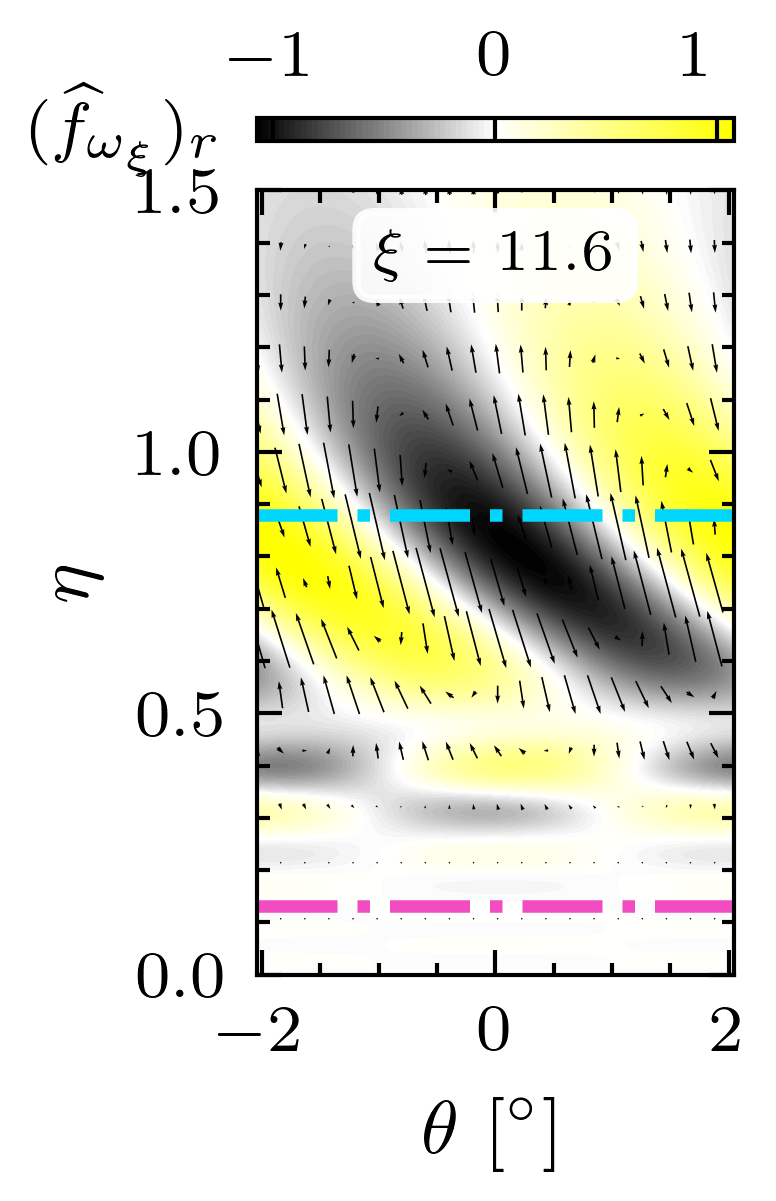}
\caption{Forcing}
\label{subfig:Results/azimuthal_entropy_layer_f35_m88_forcing_11}
\end{subfigure}
\begin{subfigure}{0.245\textwidth}
\captionsetup{width = \linewidth}
\includegraphics[]{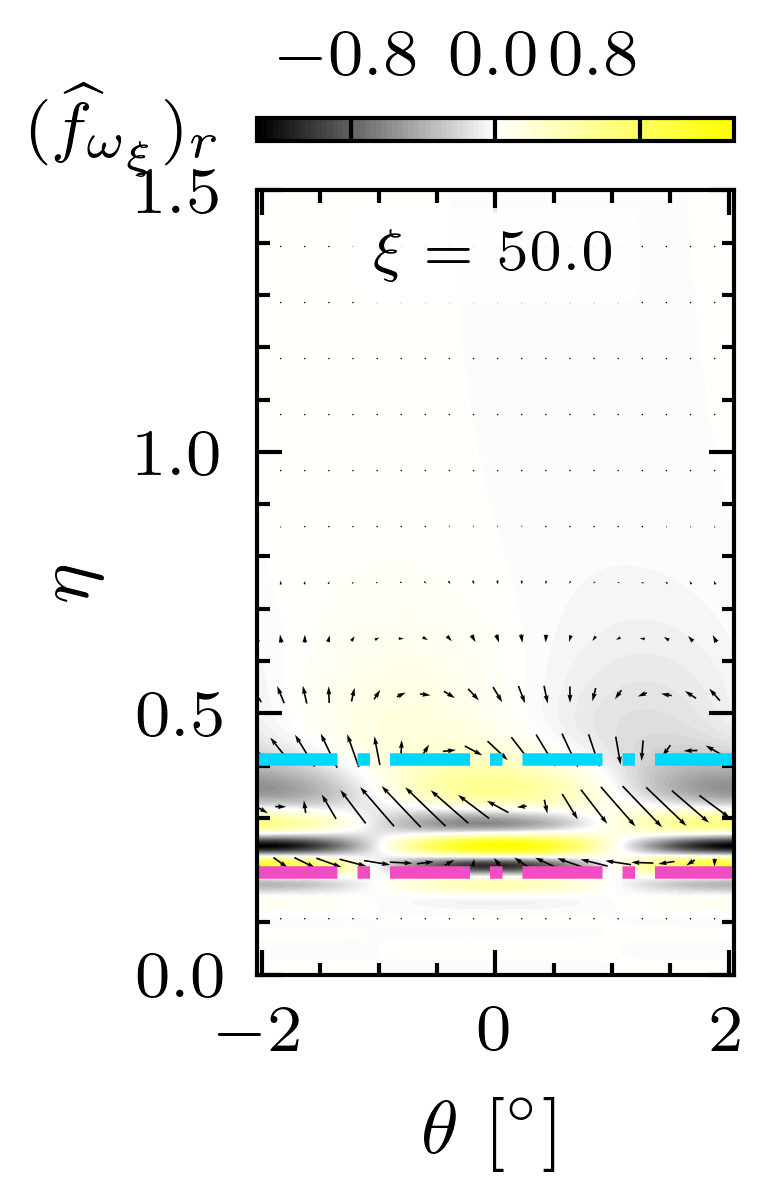}
\caption{Forcing}
\label{subfig:Results/azimuthal_entropy_layer_f35_m88_forcing_50}
\end{subfigure}
\begin{subfigure}{0.245\textwidth} 
\captionsetup{width = \linewidth}
\includegraphics[]{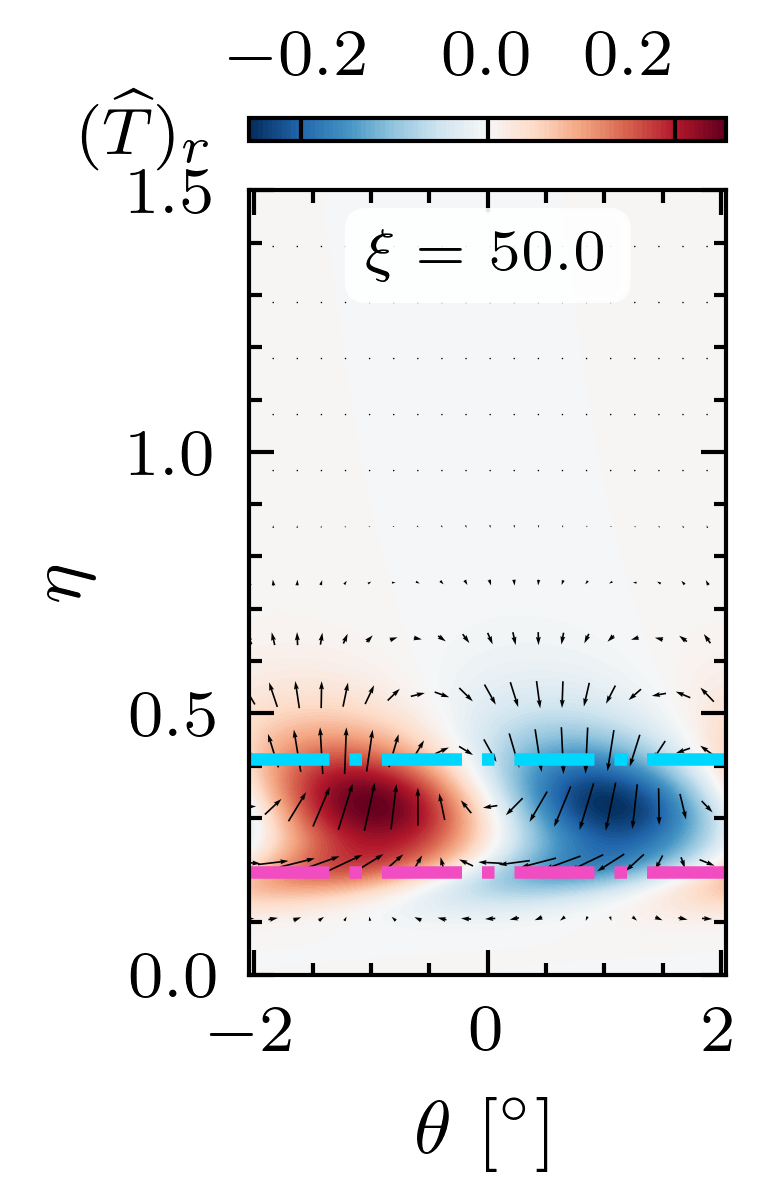}
\caption{Response}
\label{subfig:Results/azimuthal_entropy_layer_f35_m88_response_50}
\end{subfigure}
\begin{subfigure}{0.245\textwidth} 
\captionsetup{width = \linewidth}
\includegraphics[]{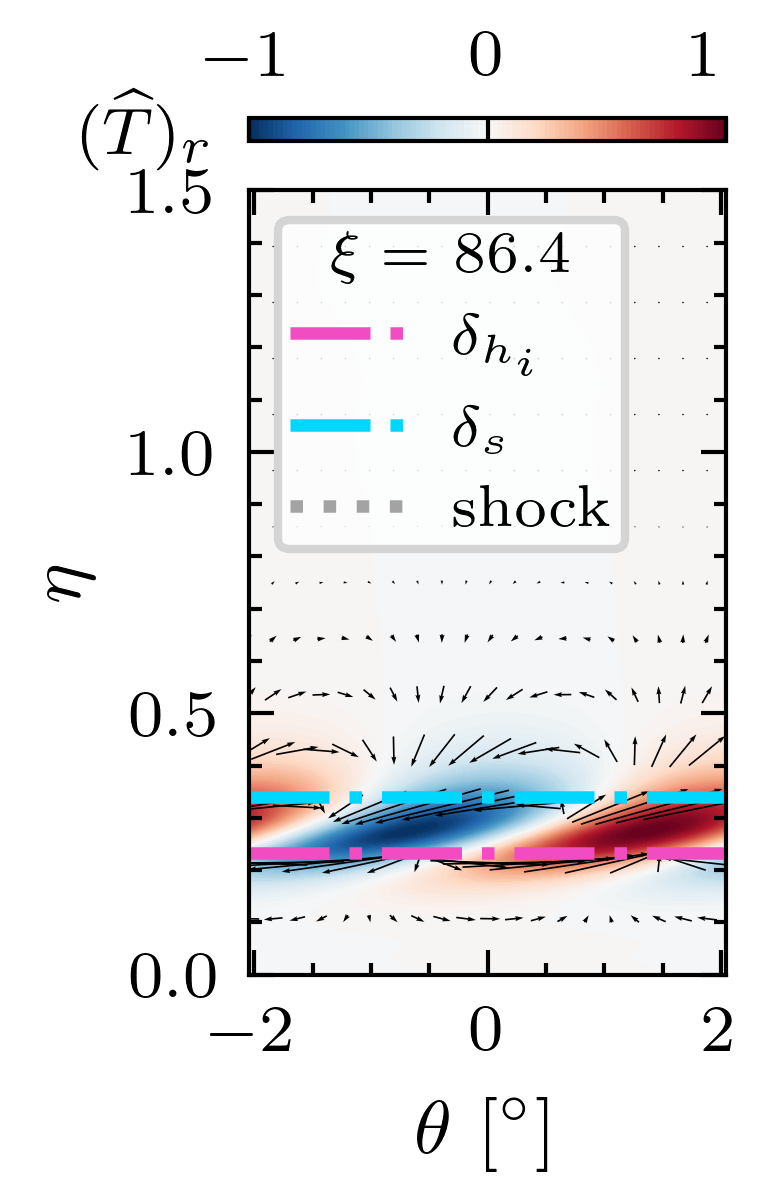}
\caption{Response}
\label{subfig:Results/azimuthal_entropy_layer_f35_m88_response_86}
\end{subfigure}
\captionsetup{width= 0.93\linewidth}
\caption[]{Real part of the forcing velocity vector field with associated streamwise vorticity \textbf{(left)} \& real part of the response velocity and temperature \textbf{(right)} for the entropy layer mode}
\label{fig:Results/azimuthal_entropy_layer_f35_m88_response}
\end{figure}

\begin{figure}
\captionsetup[subfigure]{oneside, margin={0.5cm,0cm}}
\centering
\begin{subfigure}{0.245\textwidth}
    \includegraphics[]{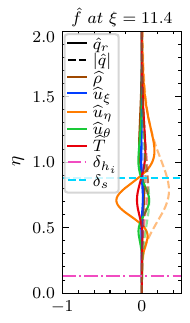}
    \captionsetup{width = \linewidth}
    \caption{Forcing}
    \label{fig:Results/slice_entropy_layer_f35_m88_forcing_xi_Rn11}
\end{subfigure}
\begin{subfigure}{0.245\textwidth}
    \includegraphics[]{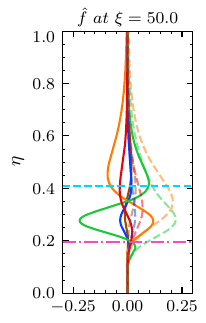}
    \captionsetup{width = \linewidth}
    \caption{Forcing}
    \label{fig:Results/slice_entropy_layer_f35_m88_forcing_xi_Rn50}
\end{subfigure}
\begin{subfigure}{0.245\textwidth}
    \includegraphics[]{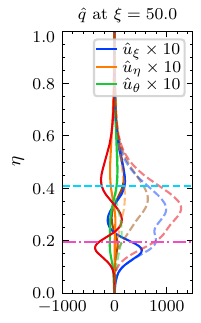}
    \captionsetup{width = \linewidth}
    \caption{Response}
    \label{fig:Results/slice_entropy_layer_f35_m88_response_xi_Rn50}
\end{subfigure}
\begin{subfigure}{0.245\textwidth}
    \includegraphics[]{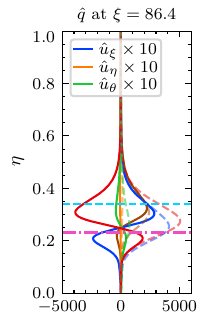}
    \captionsetup{width = \linewidth}
    \caption{Response}
    \label{fig:Results/slice_entropy_layer_f35_m88_response_xi_Rn86}
\end{subfigure}
\caption{Non-dimensionalised  local profiles of the entropy layer mode corresponding to the maximum forcing, an intermediate position and the maximum response curvilinear abscissa respectively}
\label{fig:Resultats/slice_entropy_layer_mode}
\end{figure}

To understand how the mode is structured, Figures \ref{fig:Resultats/domain_entropy_layer_f35_m88_Vf_T} and \ref{fig:Resultats/domain_entropy_layer_f35_m88_Wf_T} show $\left(\xi, \eta\right)$ streamwise planes for both wall-normal and azimuthal velocity forcing as well as the temperature response, which are the dominant primitive variables in the energy analysis. 
Figure \ref{fig:Results/azimuthal_entropy_layer_f35_m88_response} presents $\left(\theta, \eta\right)$ cross-sections showing, respectively, the forcing velocity field and the associated streamwise vorticity and the response velocity and temperature fields. 
For completeness, Figure \ref{fig:Resultats/slice_entropy_layer_mode} shows the forcing and response profiles at the same locations. 
A first consideration to note is that this mode has most of its forcing and response located far from the boundary layer, at the upper extremity of the entropy layer. 
However, when the entropy layer is less radially separated from the boundary layer, the forcing and the response also have some support at the top of the boundary layer.
The forcing is primarily composed of wall-normal velocity whose radial support extends beyond the entropy layer limit, according to the definition given in section \ref{section:Baseflow_charateristics}. 
The forcing at these upstream positions is predominantly comprised of wall-normal velocity fluctuations of alternating sign (in $\theta$ and $\xi$). 
The associated streamwise vorticity forms bands with large azimuthal inclination. 
Downstream, after $\xi = 40$, azimuthal velocity forcing becomes more important and the forcing structure takes the form of a streamwise vortex, see also Figure \ref{fig:Results/slice_entropy_layer_f35_m88_forcing_xi_Rn50}.
The response emerges downstream as oscillating temperature in both the streamwise and azimuthal directions, with maximum support between the boundary and entropy layer edges.
The forcing velocity structure is initially inclined against the velocity gradient, whereas the response structure becomes more and more inclined with the velocity gradient as it moves downstream.
This gradual rotation of the structures from upstream to downstream is reminiscent of an Orr-like mechanism, it occurs here in the entropy layer rather than the boundary layer. 
The underlying cause may lie in the baseflow: between the boundary and entropy layer edges, the streamwise velocity continues to increase, forming a region of shear that convects the mode downstream with a velocity that varies with $\eta$ within the entropy layer. 
In the azimuthal direction, the mode rotates clockwise, when $m>0$, from upstream to downstream such that high temperature response zones are associated with upwards wall-normal velocity and inversely for cool temperature response zones such that high-temperature flow is lifted from the boundary layer into the entropy layer whereas low-temperature flow is drawn into the entropy layer from above.

\subsection{Energy budget analysis of the entropy layer mode at  \texorpdfstring{$f^* =$}~ 35 kHz, m = 88}
\begin{figure}
\centering
\begin{subfigure}[t]{0.32\textwidth} \includegraphics[]{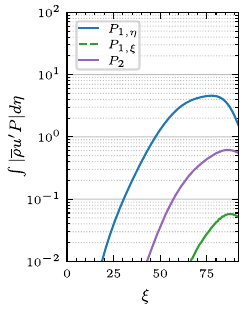}
	\caption{}
	\label{fig:Resultats/KE_production_mode_f35000_m88}
\end{subfigure}
\begin{subfigure}[t]{0.32\textwidth} \includegraphics[]{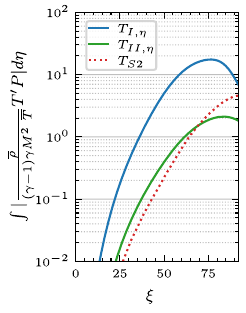}
	\caption{}
	\label{fig:Resultats/T_production_mode_f35000_m88}
\end{subfigure}
\caption{Wall-normal integrated dominant production terms for the linearised kinetic energy equation (left, \ref{fig:Resultats/KE_production_mode_f35000_m88}) and temperature equation (right, \ref{fig:Resultats/T_production_mode_f35000_m88}) for the entropy layer mode}
\end{figure}

To get a full picture of the mechanisms associated with this mode, an energy budget analysis is performed in a similar manner to that undertaken for the optimal streak mode. Figure \ref{fig:Resultats/KE_production_mode_f35000_m88} shows the most significant terms of kinetic energy production. 
The shear production term, is dominant everywhere and peaks around $\xi = 80$ shortly before the response Chu energy density maximum. 
Before it reaches the maximum production, there are two regions of exponential slope that can be identified. The first up to $\xi/Rn = 60$ where the slope is most important and a second after and up to the maximum at $\xi = 80$ where the slope is smaller. 
The beginning of this second zone, where the production is most important, matches the point at which the azimuthal velocity forcing becomes more important than the wall-normal velocity forcing, see Figure \ref{subfig:Resultats/E_chu_entropy_layer_momentum}.

To elucidate how the T energy grows, Figure \ref{fig:Resultats/T_production_mode_f35000_m88} shows the different terms contributing to the T energy growth. 
It is the $\text{T}_{I, \eta}$ term, representing the transport of baseflow temperature by wall-normal velocity fluctuations, which dominates .
The velocity dilatation is found not to play an important role in the growth of entropy layer modes. 
Though less pronounced than for the kinetic energy production, there are also two regions in the slope of the $\text{T}_{I, \eta}$ term. 
The first region up to $\xi = 60$, were the $\text{T}_{I, \eta}$ increases the most. 
Then the second region, between $\xi = 60$ up to the peak growth at $\xi = 80$, where the $\text{T}_{I, \eta}$ energy levels out.

These different elements suggest the following mechanism : entropy layer modes are  produced by the combined action of velocity fluctuations and the baseflow temperature gradient, moving flow from regions of high to low energy. 
And, to a lesser measure, the action of the velocity fluctuations does the same to the momentum in the entropy layer. 

\subsection{Entropy layer modes characteristics}

\begin{figure}
\centering
\begin{subfigure}[t]{0.3\textwidth} 
\centering 
\includegraphics[]{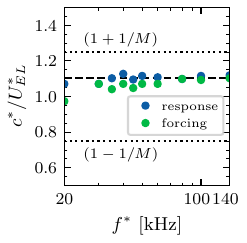}
\captionsetup{width=\linewidth}
    \caption{Phase velocity of the entropy layer modes}
    \label{fig:Resultats/phase_velocity_entropy_layer_modes}
\end{subfigure}%
\begin{subfigure}[t]{0.45\textwidth} 
\centering 
\includegraphics[]{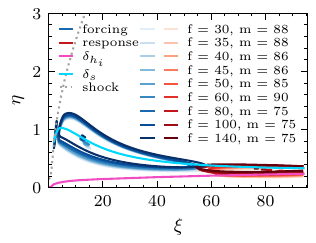}
\captionsetup{width=\linewidth}
\caption[]{Isocontour of the magnitude forcing $\vert \hat{f}_{{\Vel}_{\eta}}\vert/ max = 0.5  $ and response  $\vert \widehat{\Vel}_{\xi} \vert /max = 0.5$ for entropy layer modes with $f^* \in [30, ~140]~$ \SI{}{\kHz}}
\label{fig:Resultats/similarity_2D_shape_forcing_and_response}
\end{subfigure}%
\begin{subfigure}[t]{0.25\textwidth}
\includegraphics[width=\linewidth]{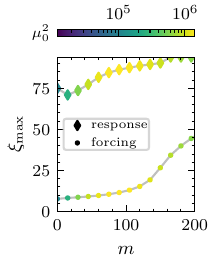}
\captionsetup{width=\linewidth}
\caption{Position of maximum forcing and response Chu energy modes at $f^* = \SI{40}{\kHz}$}
\label{subfig:Resultats/max_xi_forcing_response_40kHz_vs_m}
\end{subfigure}%
\caption{}
\label{fig:Resultats/max_xi}
\end{figure}

We now consider the whole family of HAWEL modes with extensive support of both forcing and response in the entropy layer exists for a wide range of frequencies: from $20$ to $140 ~\SI{}{\kHz}$ and azimuthal wavenumber between $75$ and $90$. 
The phase speed of the response modes, given by $c = \w/\alpha$ where $\alpha$ is the streamwise (angular) wavenumber, is plotted in Figure \ref{fig:Resultats/phase_velocity_entropy_layer_modes}. 
This was computed using the streamwise wavenumber found by taking the dominant term of the spatial Fourier transform of $\hat{\q}$ on a wall-parallel line at $\eta/R_n = 0.3$. 
The resulting phase speed is found to be constant for the response modes at the optimal azimuthal wavenumber (following the black dots in Figure \ref{subfig:Resultats/carte_gain_opt_large_iso_Rn5}) and is $1.09$ times faster than the local flow velocity at the edge of the entropy layer at end of the domain.
A similar method was used on the forcing modes and the phase velocity was found to be slightly lower in all cases.

The streamwise position of the maximum response $E_\eta$ is independent of the frequency of the entropy layer mode considered. 
Additionally, the response and forcing outline in the $(\xi, \eta)$ plane, shown in Figure \ref{fig:Resultats/similarity_2D_shape_forcing_and_response}, is also independent of frequency for the optimal entropy layer modes, indicating a self-similarity of the entropy layer modes as function of frequency and a scaling that allows collapse of the mode shapes.
However, when the azimuthal wavenumber differs significantly from the optimum, the streamwise position of peak forcing and response energy changes, see Figure \ref{subfig:Resultats/max_xi_forcing_response_40kHz_vs_m} where the optimal forcing and response positions are provided and coloured by the optimal gain value. 
The streamwise position of maximum Chu energy density is here plotted for modes at \SI{40}{\kHz} as function of $\m$. 
The position of the response continuously moves downstream as $m$ increases in the range $m \in [40, 200]$ corresponding to entropy layer modes. 
This suggests that there is a characteristic length scale associated with the baseflow that defines the range of wavenumbers and frequencies of the entropy-layer modes.
The forcing also follows the downstream trend of the peak with $m$, however the peak amplitude remains upstream until $m < 120$, then quickly move downstream for higher wavenumbers.
This indicates that the forcing process primarily favours forcing where the radial separation between the boundary and entropy layer is high. 
Combined with the observation of a self-similarity of the mode organisation with frequency, this suggest that the characteristic time and length scales of these modes may be related to the height difference $\delta_s-\delta_{h_i}$ or the angles $\theta$ formed by the entropy and boundary layer $\theta \left( \delta_s \right) - \theta \left( \delta_{h_i} \right)$.
This scale could be a key parameter underpinning the receptivity of these modes.

\section{Wall temperature effect : the adiabatic blunt cone}
\label{section:wall_temp_effects}

Past research on hypersonic boundary layer instabilities \citep{mack_stability_1987} has shown a strong dependence of the classical Mack's 1st- and 2nd-modes amplification on the wall temperature to recovery temperature ratio (i.e. the adiabatic wall temperature). 
During hypersonic flight, the wall temperature can strongly vary depending on the material nature and the type of trajectories (reentry, cruise). 
Therefore the wall temperatures can evolve between cold to warm conditions motivating the study of the effects of the wall temperature to recovery temperature ratio.
Hence, we now extend the analysis to a similar blunt cone configuration but with an adiabatic wall condition to compare it  with the cold isothermal wall studied in the previous sections.

\subsection{Resolvent gain maps}
\begin{figure}
\centering
\begin{subfigure}[t]{0.49\textwidth} \includegraphics[]{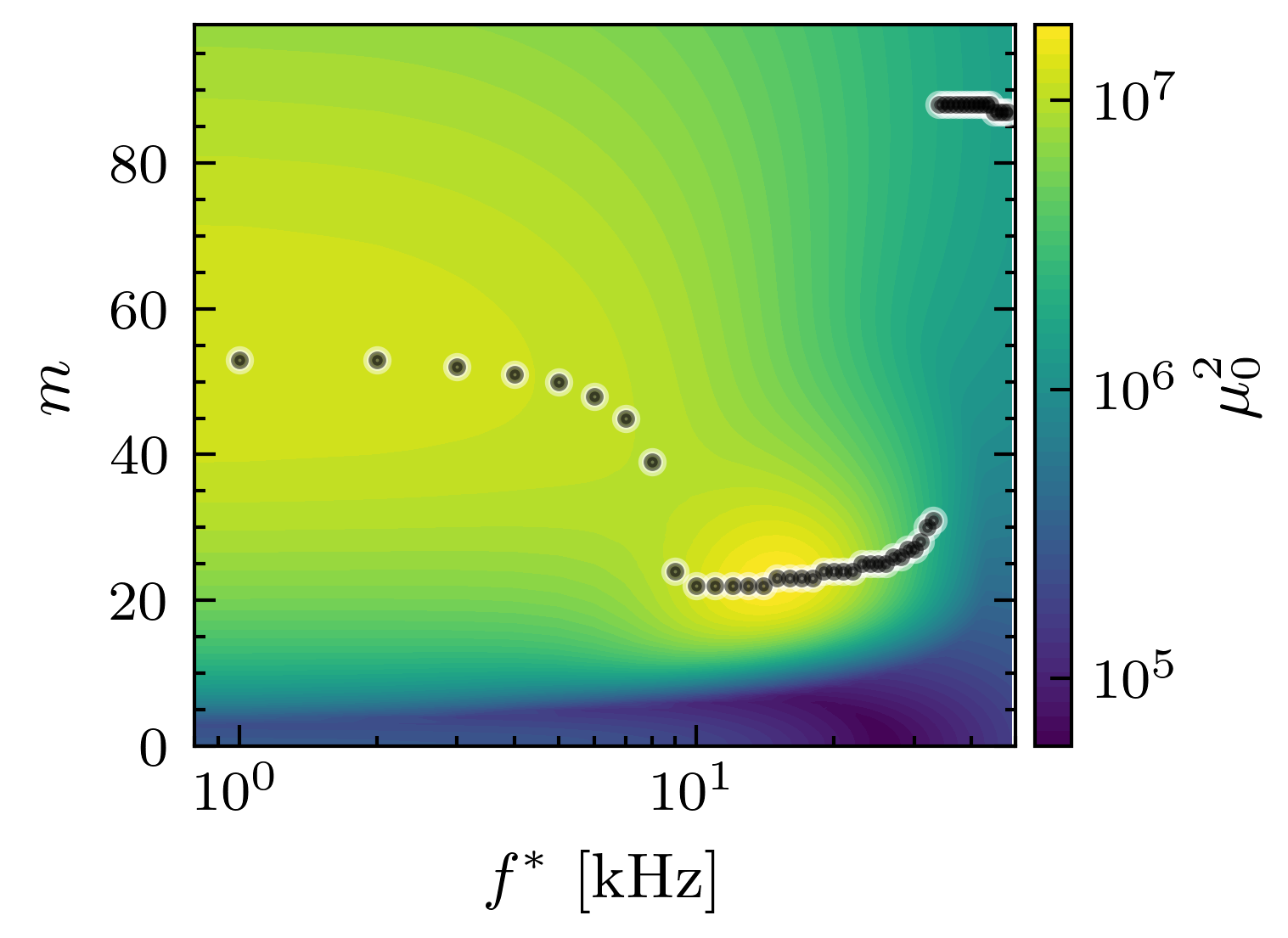}
    \caption{Adiabatic optimal gain}
    \label{subfig:Resultats/carte_gain_opt_zoom_adia}
\end{subfigure}
\begin{subfigure}[t]{0.49\textwidth} \includegraphics[]{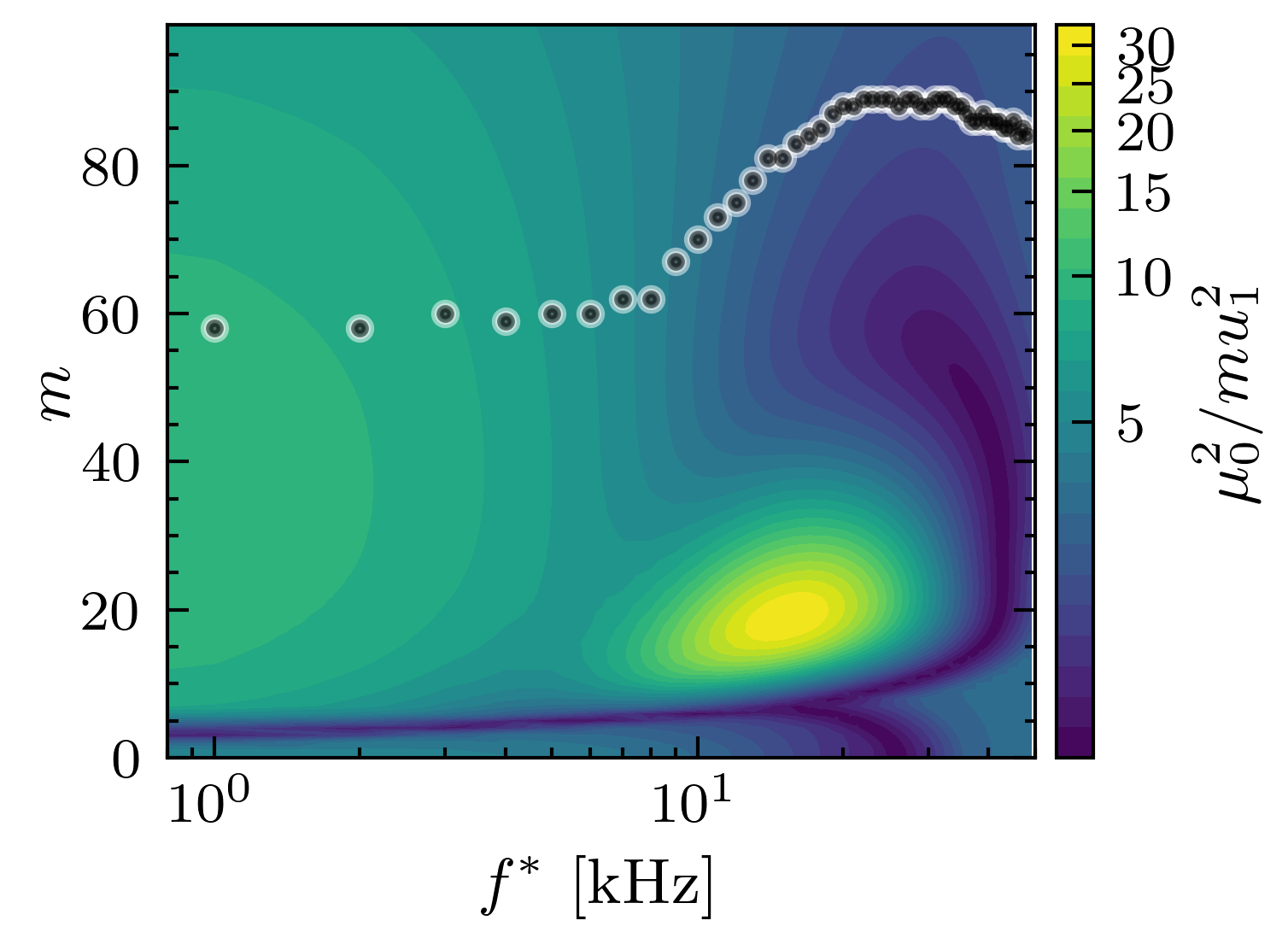}
    \caption{Gain separation} \label{subfig:Resultats/gain_seperation_map_Rn5_adia_Rn_adim_corrected_zoom_3044_469}
\end{subfigure}
\begin{subfigure}[t]{0.49\textwidth} \includegraphics[]{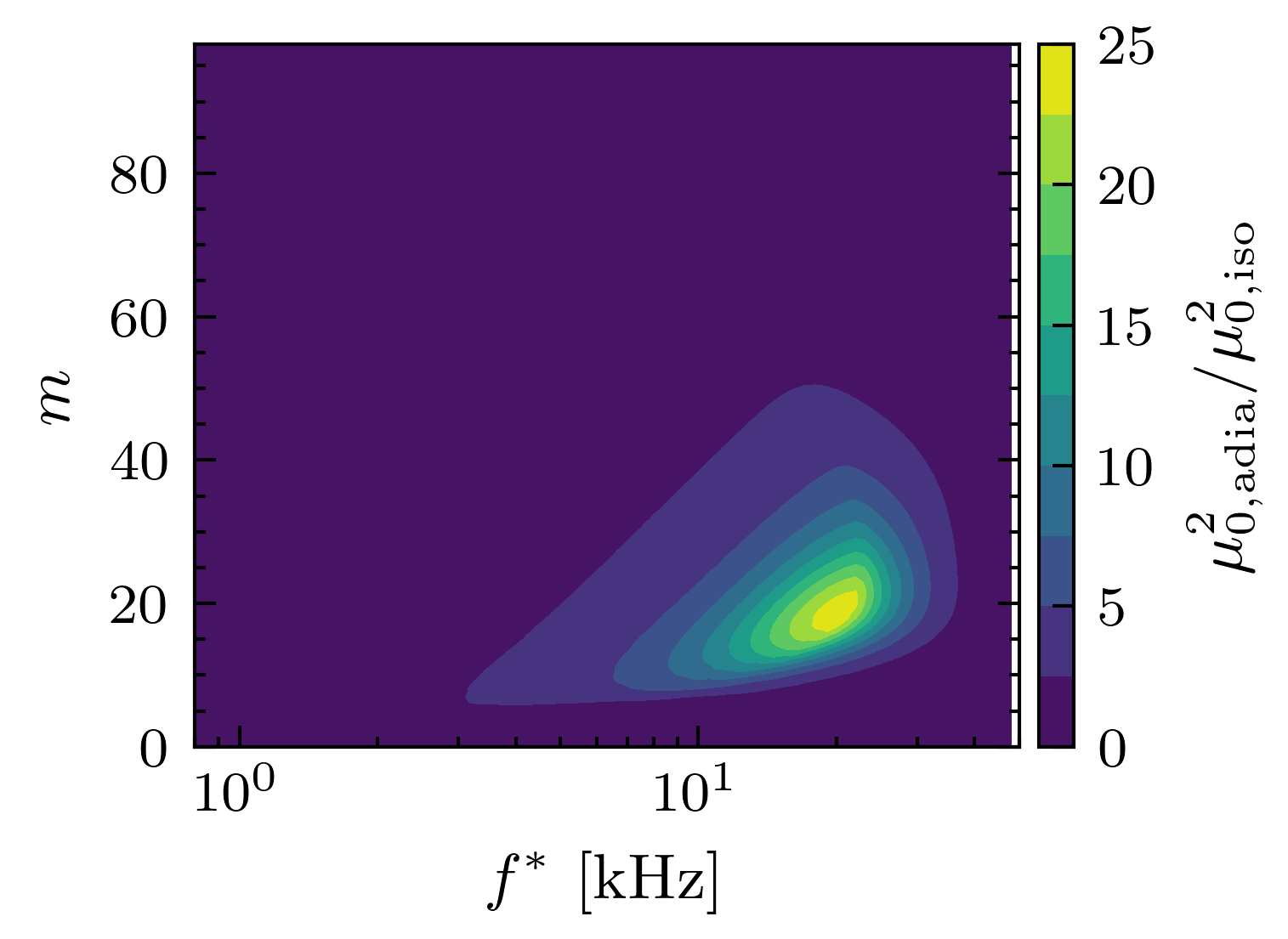}
    \caption{Adiabatic and isothermal gains ratio} \label{subfig:Resultats/gain_map_Rn5_Rn_adim_iso_vs_adim_zoom}
\end{subfigure}
\caption{Adiabatic gain maps in the $(f^*, m)$ plane. 
\blackdotwwhite ~ most amplified mode at given $f^*$}
\label{fig:Resultats/carte_gain_adia}
\end{figure}

The optimal gain map for the adiabatic wall is shown in Figure~\ref{fig:Resultats/carte_gain_adia}. In contrast to the isothermal case, the most amplified mode is no longer the steady streak mode, but rather the first Mack mode, occurring at \SI{15}{\kHz} and azimuthal wavenumber $m = 23$. 
Such an amplification is consistent with the previously documented growth of first-mode instability for adiabatic to warm wall conditions \citep{mack_stability_1987}.
This mode exhibits a gain of $\mu_0^2=1.48 \times 10^7$ \SI{}{\square\s}, slightly exceeding the gain of $\mu_0^2=1.20 \times 10^7$ \SI{}{\square\s}  observed for the isothermal streak mode but is greatly attenuated compared to the sharp case :  $\mu_0^2\approx10^{11}$ \SI{}{\square\s} for the $1^{st}$ mode and  $\mu_0^2\approx10^{13}$ \SI{}{\square\s} for the $2^{nd}$ mode. 
The significant gain separation between leading and first sub-optimal modes is consistent with a modal amplification mechanism.
At low frequency, the most amplified streak mode in the adiabatic case is the steady mode at $m = 54$, with a gain of $1.28 \times 10^7$ \SI{}{\square\s}. 
Beyond \SI{35}{\kHz}, the gain associated with the first Mack mode diminishes. 

In this higher frequency range, a family of entropy layer modes re-emerges, displaying characteristics similar to those observed in the isothermal wall case. 
The gain and gain separation of the adiabatic entropy layer modes are similar to the isothermal case, indicating that these modes are not very sensitive to the wall condition.
The shift in dominant amplification mechanisms between the isothermal and adiabatic wall conditions has practical implications for transition modelling on flight vehicles, for which the wall temperature may evolve between the colder isothermal wall and adiabatic wall cases.

\subsection{Adiabatic blunt cone optimal \texorpdfstring{$1^{st}$}{1st} Mack mode}

\begin{figure}
\centering
\includegraphics[width = \linewidth]{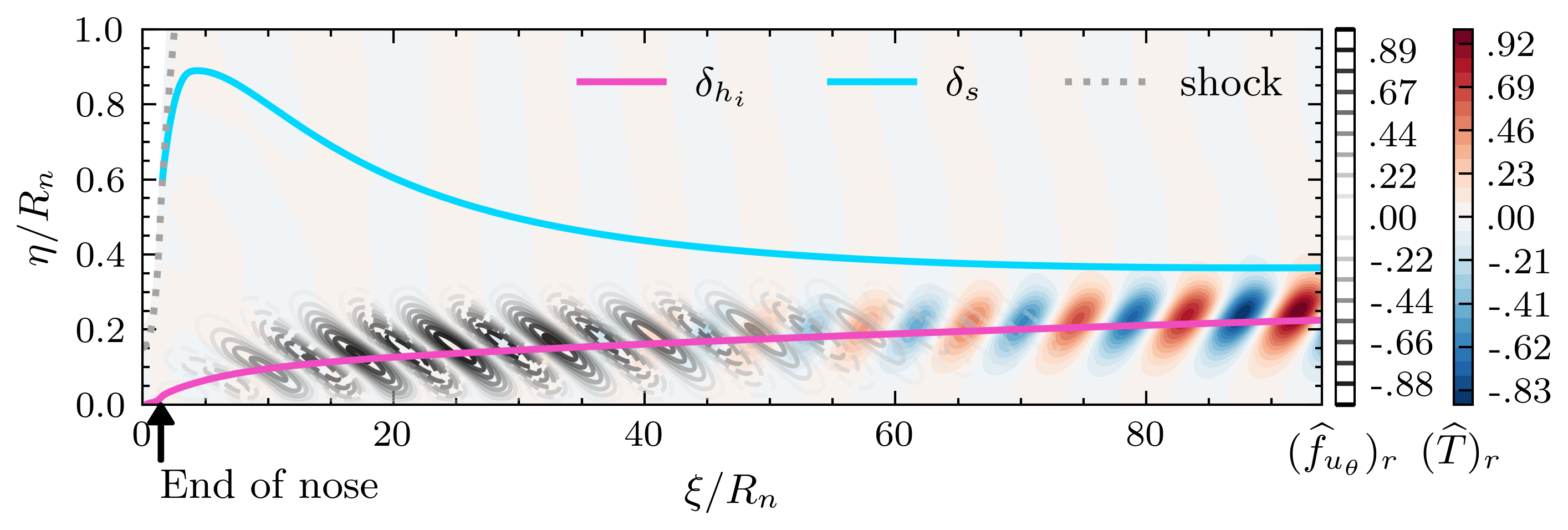}
\caption[]{Global shape of the real parts of the azimuthal velocity forcing and temperature response of the adiabatic $1^{st}$ Mack mode. Normalised by their respective maximum absolute value}
\label{fig:Resultats/domain_1st_Mack_mode_f15_m23_Wf_T}
\end{figure}

The most amplified adiabatic mode is the $1^{st}$ Mack mode, which displays similar characteristics to the canonical sharp cone $1^{st}$ Mack mode. 
A $(\xi, \eta)$ streamwise
plane of the mode is presented in Figure \ref{fig:Resultats/domain_1st_Mack_mode_f15_m23_Wf_T}.
The forcing and response is primarily located on the boundary layer limit.
The optimal $1^{st}$ Mack mode uses the Orr mechanism to maximise the initial perturbation energy. 
This is apparent in the receptivity structure that forms alternating bands inclined against the flow velocity gradient, characteristic of an Orr mechanism. 
Then the first-mode follows a modal exponential amplification as it evolves further downstream.

The presence of bluntness and the induced boundary layer causes the maximum gain of the $1^{st}$ Mack mode to be greatly reduced in comparison with the flow over an equivalent sharp cone. 
However, the entropy layer does not modify the behaviour of the $1^{st}$ Mack mode, and it is quite similar to that on a flat plate \citep{bugeat_3d_2019}.

\section{Conclusion}
\label{section:Conclusion}

Transition over hypersonic moderately blunt bodies is not well understood and past research did not provide a comprehensive review of the instabilities and receptivity existing in such flows.
The bluntness induces a detached shock which significantly alters the baseflow with respect to the sharp case, modifying the linear stability mechanisms responsible for transition.
To tackle this problem, resolvent analysis was used for a $7^\circ$ half-angle moderately blunt cone at Mach 6 with zero angle of attack and Reynolds number $\Re_{R_n} = 90000$ in order to characterise, in a comprehensive manner, the linear amplification mechanisms that may arise for a large range of frequencies and wavenumbers.
A general result is that at this intermediate Reynolds number, all Mack modes are strongly damped, especially the second mode, making the path to transition unclear. 
Resolvent analysis shows that in the isothermal wall case, streak modes with azimuthal wavenumber $m = 57$ are the most amplified followed by a family of entropy layer modes at high azimuthal wavenumber (designated as HAWEL mode) with an amplification peak at $m\approx 88$. 
A last family of modes with a signature in the entropy layer, distinct from the HAWEL modes, has been identified at lower azimutal wavenumbers ($m<21$) (designated as LAWEL mode), and have a much smaller gain compared to other modes.
One of the main result of this study lies in a first detailed description of these HAWEL optimal modes that are the leading amplification mechanisms for a large range of frequencies and wavenumbers. 
This descriptions is particularly useful as past hypersonic wind-tunnel experiments consistently displayed entropy-layer elongated fluctuations (wisps) in high-speed Schlieren imagery \citep{marineau_mach_2014, jewell_boundary-layer_2017, grossir_influence_2019, ceruzzi_experimental_2024}. 
The key findings related to these instabilities is that they represent a family of modes with : a weak amplification (6 to 20 times less than steady streaks), a clear localisation of the optimal forcing in the entropy-layer far above the boundary-layer and an optimal response located in the entropy layer just above the boundary-layer after its starts merging with the entropy-layer.
These modes are most susceptible to wall-normal velocity disturbances upstream which evolve downstream into alternating streamwise vorticity.
The response on the other hand is most important on temperature fluctuations and an energy analysis shows that it is the conjoint action of wall-normal velocity fluctuations and wall-normal gradients of baseflow temperature and velocity which extract energy from the baseflow to the fluctuations.

Stationary streaks remain the most amplified mechanism, and may be expected to appear and participate in the transition process. 
Bluntness causes the receptivity region of streaks to move into the entropy layer compared to the sharp reference. The  blunt optimal streak mode is most receptive to radial pumping before gradually entering the boundary layer, where the more usual streamwise vorticity appears in the forcing mode. 

It is interesting to note that in both the HAWEL and streak-mode scenarios, the entropy layer plays a key role in the receptivity.
Both of these modes are susceptible to wall-normal velocity forcing upstream in the entropy layer before becoming more susceptible to a longitudinal vortex structure forcing further downstream.

Modification of the baseflow by steady streaks changes the growth rate of modes \citep{vaughan_stability_2011, caillaud_effect_2025}. 
Studying the secondary stability problem would provide insight on possible transition scenarios. 
Furthermore, an interesting perspective would be to analyse the secondary stability problem of streaks associated with entropy layer modes. 

Regarding the effect of wall conditions on the linear amplification mechanisms, this being important in understanding differences between real atmospheric re-entry and windtunnel testing, our study shows that in the case of an adiabatic wall condition, the $1^{st}$ Mack mode becomes the most amplified mode in the $(f^*, m)$ plane, followed by streaks and entropy layer modes at high azimuthal wavenumber. 
This further suggests that in moderately blunt regimes, the transition process may involve a competition between streaks, $1^{st}$ Mack mode and entropy layer modes. 

While the resolvent maps out all possible linear mechanisms and gives insight into optimal receptivity, it does not take into account the space of physical perturbations that are available to force the flow. 
Revisiting these findings with the physically realisable resolvent analysis developed by \cite{kamal_global_2023} will provide further clarification of how the amplification mechanisms identified and discussed in this paper are activated by atmospheric perturbations.
Alternatively, with knowledge of the environmental disturbance, projection of these  on the optimal resolvent basis would enable the calculation of a weighted optimal gain map, possibly highlighting new mechanisms.

Finally, the transition process is not only a linear problem. Studying multimode transition scenarios will also be key in understanding the blunt cone transition process. 

\backsection[Declaration of interests]{The authors report no conflict of interest}

\appendix

\section{}\label{appA}
\subsection{Resolvent mesh convergence}

\begin{figure}
\centering
\begin{subfigure}[T]{0.29\textwidth} 
\centering
\captionsetup{width = \linewidth}
\includegraphics[]{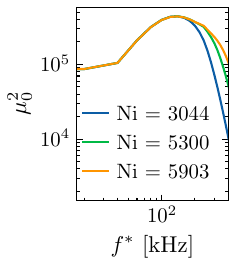}
\caption[]{Optimal gain for different meshes at $m=0$.}
\label{subfig:AppendixA/mesh_convergence_gains}
\end{subfigure}
\begin{subfigure}[T]{0.4\textwidth} 
\centering
\captionsetup{width = \linewidth}
\includegraphics[]{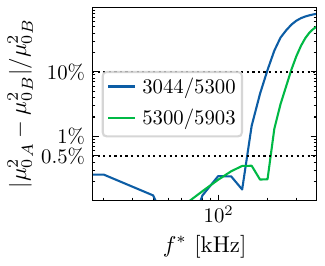}
\caption[]{Relative error of mesh A compared to the finer mesh B}
\label{sufig:AppendixA/mesh_convergence_relative_error}
\end{subfigure}
\begin{subfigure}[T]{0.29\textwidth} 
\centering
\captionsetup{width = \linewidth}
\includegraphics[]{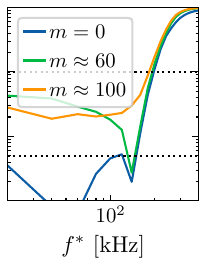}
\caption[]{Relative error between meshes with 3044 and 5903 points for different $m$.}
\label{sufig:AppendixA/mesh_convergence_gains_different_m}
\end{subfigure}
\caption{}
\label{fig:AppendixA/mesh_convergence}
\end{figure}

The mesh convergence for the resolvent analysis is based on the convergence of the optimal gain. 
In figure \ref{fig:AppendixA/mesh_convergence}, the optimal gain is plotted for different meshes with increasing streamwise resolution at $m = 0$. 
Only the streamwise resolution convergence is presented as the resolved wall-normal length-scale is much greater than that of the observed structures.
Very good agreement is obtained for the roughest mesh with $Ni = 3044$ points in the streamwise direction until \SI{150}{\kHz}, where the relative error to the finer meshes crosses the $0.5\%$ threshold. 
This mesh was used for all the analysis. 
The relative error increases to $10\%$ above  \SI{200}{\kHz} because the gain is underestimated, nonetheless, the trend is preserved and the mode structure is largely preserved. 

\begin{figure}
\centering
\begin{subfigure}{0.245\textwidth}
    \centering
    \includegraphics[]{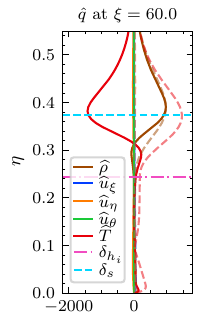}
    \captionsetup{width = \linewidth}
    \caption{Response}
    \label{fig:slice_entropy_layer_f260_m0_Response_xi_Rn60}
\end{subfigure}
\begin{subfigure}{0.245\textwidth}
    \centering
    \includegraphics[]{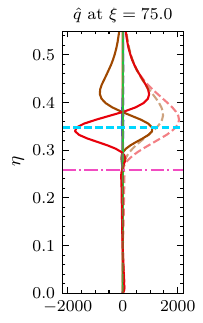}
    \captionsetup{width = \linewidth}
    \caption{Response}
    \label{fig:slice_entropy_layer_f260_m0_Response_xi_Rn75}
\end{subfigure}
\caption{Optimal response of the mode  $(f^*, m) = (260~ \text{kHz}, 0)$ computed with the high resolution mesh Ni = 5903 at two streamwise positions where the response is strong}
\label{fig:Resultats/slice_entropy_layer_f260_m0_response}
\end{figure}

The highest frequency necessary to resolve is that of the $2^{nd}$ Mack mode. For the sharp case the $2^{nd}$ mack mode becomes optimal at \SI{300}{\kHz} and peaks at \SI{330}{\kHz}. Based on the relation $f_{2^{nd}}^* \propto U_e/\delta$ the $2^{nd}$ mode is expected to range from $240$ to \SI{260}{\kHz} in the blunt case due to the thicker boundary layer. 
At these frequencies, the mesh with $Ni = 3044$ points is not well resolved but the $5300$ mesh is with less than 10\% error.
There is no substantial increase in optimal gain as would be expected with the growth of the second Mack mode.
Furthermore, looking at the profile of the mode obtained in this range of frequencies, it is clear that it does not correspond to the $2^{nd}$ mode, see Figure \ref{fig:Resultats/slice_entropy_layer_f260_m0_response}. 
The $2^{nd}$ Mack mode is truly attenuated to the point that it is no longer an optimal resolvent mode. 

\section{Profiles of suboptimal modes}
\label{Appendix:suboptimal_profiles}

To complete the discussion on suboptimal modes started in section \ref{section:suboptimal} this Appendix presents additional profiles of the modes that were discussed. 

\begin{figure}
\centering
\begin{subfigure}[t]{\textwidth} \includegraphics[]{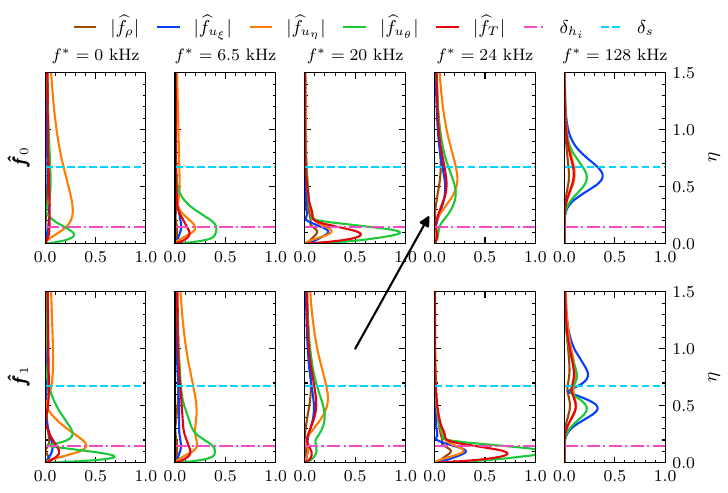}
\caption[]{Optimal forcing (top, mode 0) and $1^{st}$ suboptimal response (bottom, mode 1) at $m = 21$ at $\xi = 20$} 
\label{subfig:Resultats/sub_slice_m_forcing_21}
\end{subfigure}
\begin{subfigure}[t]{\textwidth} \includegraphics[]{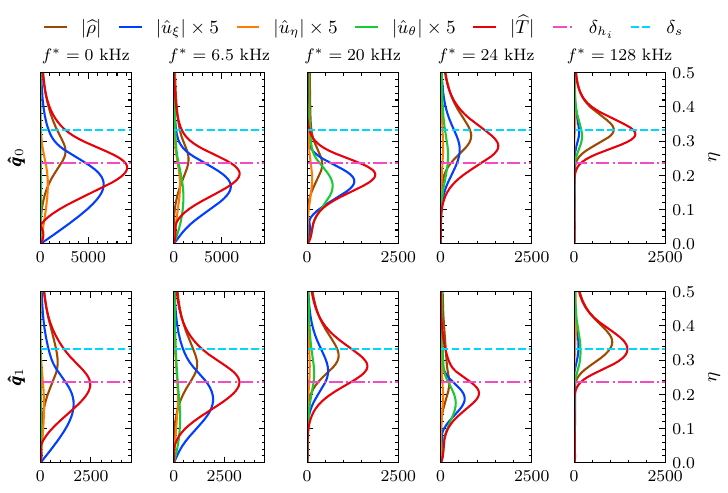}
\caption[]{Optimal response (top, mode 0) and $1^{st}$ suboptimal response (bottom, mode 1) at $m = 21$ at $\xi = 94$} 
\label{subfig:Resultats/sub_slice_m_response_21}
\end{subfigure}
\caption[]{ }
\label{fig:Resultats/sub_slice_m_21}
\end{figure}

In Figure \ref{fig:Resultats/sub_slice_m_21} are presented the two first optimal and suboptimal modes at $m = 21$ at some key frequencies. 
At $f^* = 20$ \SI{}{\kHz}, we can observe that the forcing profile is characteristic of the $1^{st}$ mode and is similar to what is observed for a flat plate \citep{bugeat_3d_2019}.
Between $f^*= 20$ and \SI{24}{\kHz}, we can observe the switch in dominant mechanisms described based on the gains in Figure \ref{subfig:Resultats/gains_and_subgains_vs_f_m21}. 
This apparent by the switch in the shape of the forcing profiles where the optimal mode forcing at $f^* = 24$ \SI{}{kHz} resembles closely to the suboptimal mode at $f^* = 20$ \SI{}{\kHz} and the characteristic shape of the $1^{st}$ mode appears in the suboptimal mode at $f^* = 24$ \SI{}{\kHz}.

\begin{figure}
    \centering
    \includegraphics[width = 0.9\linewidth]{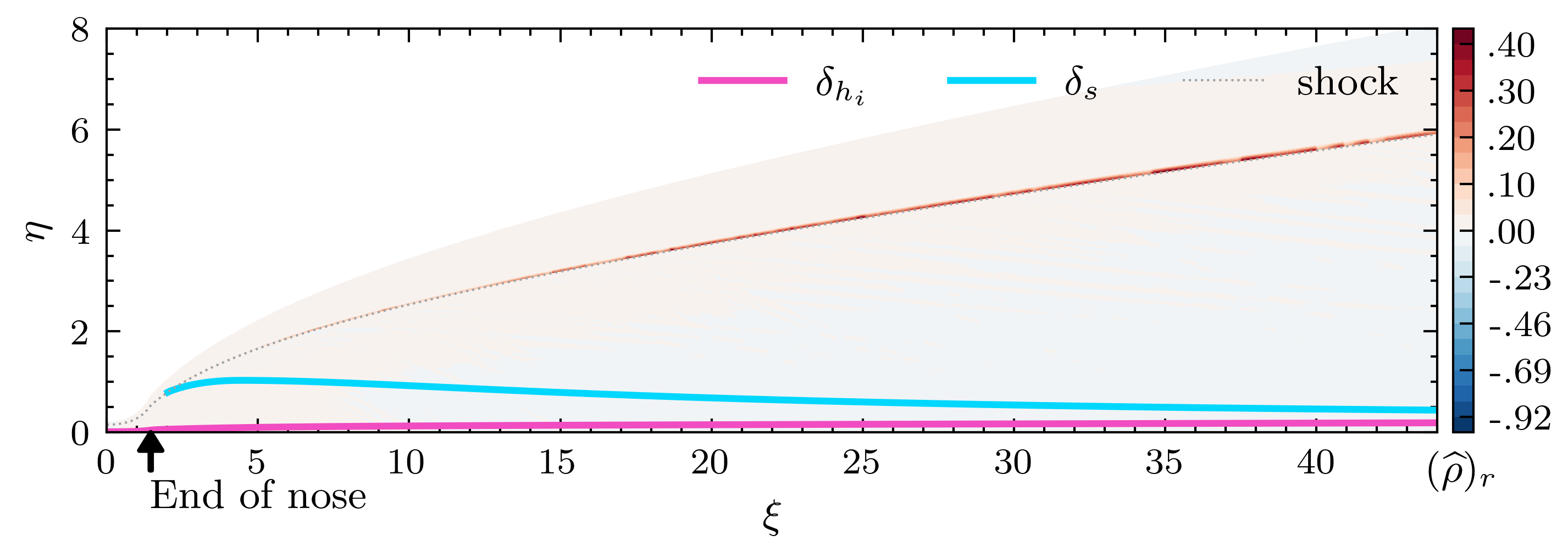}
    \caption{$(f^*, m) = (1 \text{ kHz}, 0)$ shock mode}
    \label{fig:shock_mode}
\end{figure}

In section \ref{section:suboptimal} was also mentioned the presence of axisymmetric shock modes at $m = 0$ and low frequencies. Here is an example in Figure \ref{fig:shock_mode} at $(f^*, m) = (1 ~\text{kHz}, 0)$.

\section{Linearised transport of streamwise velocity}
\label{Appendix:Linearised_eq}

The complete transport equations of velocity and temperature fluctuation used the energy budget performed in section \ref{section:OptimalStreakMode} and \ref{subsection:entropy_layer_f35_m88} are given here in equations \ref{eq:complete_vel_fluct_vel_eq} and \ref{eq:complete_temperature_fluct_trasnport_eq}. 

\begin{align}
    \partial_t \Vvec' + \left( \mean{\Vvec}\pmb{\cdot} \nabla \right) \Vvec' &= - \left( \Vvec'\pmb{\cdot} \nabla \right) \mean{\Vvec} - \frac{\Density'}{\mean{\Density}} \left( \mean{\Vvec}\pmb{\cdot} \nabla \right) \mean{\Vvec} - \frac{\nabla p'}{\mean{\Density}} + \frac{1}{\mean{\Density}} \nabla \pmb{\cdot} \fluctuation{\Ttau} + \frac{\f_{\Density \Vvec}}{\mean{\Density}}  \label{eq:complete_vel_fluct_vel_eq}\\
    \partial_t \fluctuation{\Temp} + \mean{\Vvec} \pmb{\cdot} (\nabla \fluctuation{\Temp}) &= -\Vvec' \pmb{\cdot} (\nabla \mean{T}) - \frac{\Density'}{\mean{\Density}} \mean{\Vvec} \pmb{\cdot} (\nabla \mean{T}) \nonumber \\
    & \quad - \frac{\gamma - 1}{\gamma \M_{\infty}^2} \fluctuation{\Temp} (\nabla \pmb{\cdot} \mean{\Vvec}) 
    + \frac{\gamma - 1}{\Density_{\infty} \Rey ~\mean{\Density}} (\nabla \mean{\Vvec} \colon \fluctuation{\Ttau}) 
    + \frac{\gamma}{\Pran \Rey ~\mean{\Density}} (\nabla \pmb{\cdot} (\mean{\Viscosity} \nabla \fluctuation{\Temp})) \nonumber \\
    & \quad - \frac{\gamma - 1}{\gamma \M_{\infty}^2} \mean{T} (\nabla \pmb{\cdot} \Vvec') 
    + \frac{\gamma - 1}{\Density_{\infty} \Rey ~\mean{\Density}} (\nabla \Vvec' \colon \mean{\Ttau}) + \frac{\gamma}{\Pran \Rey~\mean{\Density}} (\nabla \pmb{\cdot} (\Viscosity' \nabla \mean{T})) \nonumber \\
    & \quad + \frac{\gamma \M_{\infty}^2 (\gamma - 1)}{\mean{\Density}} \left( \normalf_{\Density E} -  \f_{\Density \Vvec} \pmb{\cdot} \mean{\Vvec} \right)    \label{eq:complete_temperature_fluct_trasnport_eq}
\end{align}

To compute the production and source terms used in the energy budget analysis, we begin by multiplying Equation \ref{eq:complete_vel_fluct_vel_eq} by $\fluctuation{\Vxi} \mathbf{e}_\xi$ and Equation \ref{eq:complete_temperature_fluct_trasnport_eq} by $\Temp'$. 
These equations are then respectively scaled by $\mean{\Density}$ and $\frac{\mean{\Density}}{(\gamma-1)\gamma M_{\infty}^2 \mean{\Temp}}$ to yield the terms that contribute to the Chu energy (see Equation \ref{eq:chu_energy}).
Finally, the wall-normal integral of the temporal and azimuthal mean is taken : $\left< \cdot \right> =  \frac{mf}{2\pi}\int_{0}^{\eta_{\max}}\int_{0}^{2\pi/m}  \int_0^{1/f} \cdot dt d\theta d\eta$, when $m \ne 0$ and $f \neq 0$, otherwise the associated mean is dropped.
The complete equations are thus,
\begin{align}
    \mean{\Density} \partial_t {\fluctuation{\Vvec}} \cdot \fluctuation{\Vxi}  \mathbf{e}_\xi+  \text{KE}_{\text{Tr}, \xi} &= \text{KE}_{\text{P1}, \xi} + \text{KE}_{\text{P2}, \xi} + \text{KE}_{\text{S}, \xi} + \text{KE}_{\text{F}, \xi} \tag{\ref{eq:KE_equation}}\\
    \frac{\mean{\Density}}{2(\gamma-1)\gamma M_{\infty}^2 \mean{\Temp}}  \partial_t {\fluctuation{\Temp}}^2 + \text{T}_{\text{Tr}} &=  \text{T}_{\text{I}} +  \text{T}_{\text{II}} + \text{T}_{\text{S1}} + \text{T}_{\text{V1}} + \text{T}_{\text{V}\mu 1} \tag{\ref{eq:TE_equation}}\\ 
    & ~+ \text{T}_{\text{S2}} + \text{T}_{\text{V2}} + \text{T}_{\text{V}\mu 2} + \text{T}_{\text{F}} \nonumber
\end{align}

\noindent where,
\begingroup
\allowdisplaybreaks
\begin{align}
    \text{KE}_{\text{Tr}, \xi}\left( \xi \right) &=  \left<  \mean{\Density} \left( \mean{\Vvec}\pmb{\cdot} \nabla \right) \fluctuation{\Vvec}  \pmb{\cdot} \fluctuation{\Vxi} \mathbf{e}_\xi \right> \\
    \text{KE}_{\text{P1}, \xi}\left( \xi \right) &=  \left< - \mean{\Density} \left( \fluctuation{\Vvec}\pmb{\cdot} \nabla \right) \mean{\Vvec}  \pmb{\cdot} \fluctuation{\Vxi} \mathbf{e}_\xi \right> \tag{\ref{eq:KE_P1}} \\
    \text{KE}_{\text{P2}, \xi}\left( \xi \right) &=  \left<  - \fluctuation{\Density} \left( \mean{\Vvec}\pmb{\cdot} \nabla \right) \mean{\Vvec}  \pmb{\cdot} \fluctuation{\Vxi} \mathbf{e}_\xi \right> \tag{\ref{eq:KE_P2}}\\
    \text{KE}_{\text{S}, \xi}\left( \xi \right) &=  \left<  - {\nabla p'}  \pmb{\cdot} \fluctuation{\Vxi} \mathbf{e}_\xi \right> \\
    \text{KE}_{\text{V}, \xi}\left( \xi \right) &=  \left<  \nabla \pmb{\cdot} \fluctuation{\Ttau} \pmb{\cdot} \fluctuation{\Vxi} \mathbf{e}_\xi \right> \\
    \text{KE}_{\text{V}, \xi}\left( \xi \right) &=  \left<   {\f_{\Density \Vvec}} \pmb{\cdot} \fluctuation{\Vxi} \mathbf{e}_\xi \right> \\
    \text{T}_{\text{T}}\left( \xi \right) &=  \left< \frac{\mean{\Density}}{(\gamma-1)\gamma M_{\infty}^2 \mean{\Temp}}  \mean{\Vvec} \pmb{\cdot} (\nabla \fluctuation{\Temp})  \fluctuation{\Temp}\right> \\
    \text{T}_{\text{I}}\left( \xi \right) &=  \left< -\frac{\mean{\Density}}{(\gamma-1)\gamma M_{\infty}^2 \mean{\Temp}}  \fluctuation{\Vvec} \pmb{\cdot} (\nabla \mean{\Temp})  \fluctuation{\Temp}\right> \tag{\ref{eq:T_P1}}\\
    \text{T}_{\text{II}}\left( \xi \right) &=  \left< -  \frac{1}{(\gamma-1)\gamma M_{\infty}^2 \mean{\Temp}} \Density' \mean{\Vvec} \pmb{\cdot} (\nabla \mean{T})  \fluctuation{\Temp}\right> \tag{\ref{eq:T_P2}}\\
    \text{T}_{\text{S1}}\left( \xi \right) &=  \left<  -\frac{\mean{\Density}}{\gamma^2 M_{\infty}^4 }  \fluctuation{\Temp} (\nabla \pmb{\cdot} \mean{\Vvec}) \fluctuation{\Temp}\right> \\
    \text{T}_{\text{V1}}\left( \xi \right) &=  \left<  \frac{1}{\gamma M_{\infty}^2 \mean{\Temp} \Density_{\infty} \Rey } (\nabla \mean{\Vvec} \colon \fluctuation{\Ttau}) \fluctuation{\Temp}\right> \\
    \text{T}_{\text{V1} \mu}\left( \xi \right) &=  \left<  \frac{1}{(\gamma-1) M_{\infty}^2 \mean{\Temp} \Pran \Rey} (\nabla \pmb{\cdot} (\mean{\Viscosity} \nabla \fluctuation{\Temp})) \fluctuation{\Temp}\right> \\
    \text{T}_{\text{S2}}\left( \xi \right) &=  \left<  -\frac{\mean{\Density}}{\gamma^2 M_{\infty}^4 }  (\nabla \pmb{\cdot} \fluctuation{\Vvec}) \fluctuation{\Temp} \right>   \tag{\ref{eq:T_S2}} \\
    \text{T}_{\text{V2}}\left( \xi \right) &=  \left<  \frac{1}{\gamma M_{\infty}^2 \mean{\Temp} \Density_{\infty} \Rey} (\nabla \fluctuation{\Vvec} \colon \mean{\Ttau}) \fluctuation{\Temp}\right> \\
    \text{T}_{\text{V2} \mu}\left( \xi \right) &=  \left<  \frac{1}{(\gamma-1) M_{\infty}^2 \mean{\Temp} \Pran \Rey } (\nabla \pmb{\cdot} (\fluctuation{\Viscosity} \nabla \mean{\Temp})) \fluctuation{\Temp}\right> \\
    \text{T}_{\text{F}}\left( \xi \right) &=  \left<  \frac{1}{\mean{\Density}} \left( \normalf_{\Density E} -  \f_{\Density \Vvec} \pmb{\cdot} \mean{\Vvec} \right) \fluctuation{\Temp}\right> \\
\end{align}.
\endgroup

\section{Radius of curvature of the baseflow}
\label{Appendix:curvature}

\begin{figure}
\centering
\includegraphics[width = \linewidth]{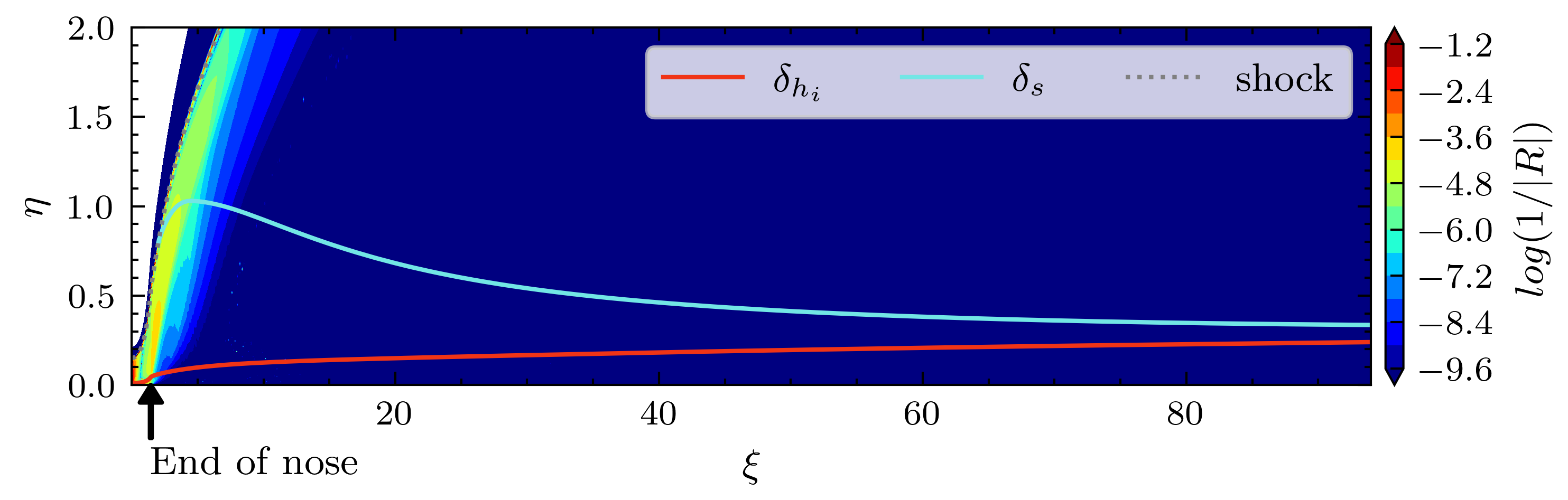}
\caption[]{Inverse of the local radius of curvature of the flow}
\label{fig:Appendix/Curvature_radius_BF}
\end{figure}

The flow curvature can be determined based on the local flow velocity by:
\begin{equation}
    R = \frac{|\Vvec|^{3}}{\U D_t\V - \V D_t\U}
\end{equation}
where $D_t$ is the material derivative. The inverse of the local radius of curvature of the flow is shown in Figure \ref{fig:Appendix/Curvature_radius_BF}. 
The maximum flow curvature is $R > 10^{10} \approx 10^{7}$\SI{}{\m}
The area of highest flow curvature is located near the stagnation point and at the start of the cone and has values of the order of $R \approx 100 \approx 0.5$ \SI{}{\mm}.

\bibliography{Bibliography/article_blunt_2024}
\bibliographystyle{./Forme/jfm}

\end{document}